\documentclass[letterpaper,twocolumn,10pt]{article}
\usepackage{usenix2019_v3}
\usepackage{changepage}
\usepackage{tikz}
\usepackage{amsmath}
\usepackage{enumitem}
\usepackage{pifont}
\usepackage{multirow}
\usepackage{caption}
\usepackage{graphicx} 
\usepackage{subcaption} 
\usepackage{amsfonts}
\usepackage{placeins}
\usepackage{filecontents}
\usepackage{tabularx}
\usepackage{booktabs} 
\usepackage{array}
\usepackage{amsmath} 
\usepackage{algorithm}
\usepackage{algorithmic}   

\begin{document}
	
	\date{}
	
	\title{\Large \bf SNAKE: A Sustainable and Multi-functional Traffic Analysis System utilizing Specialized Large-Scale Models with a Mixture of Experts Architecture}
	\author{
		{\rm Tian Qin}\\
		Southeast University
		\and
		{\rm  Guang Cheng}\\
		Southeast University
		\and
		{\rm Yuyang Zhou}\\
		Southeast University
		\and
		{\rm  Zihan Chen}\\
		Southeast University
		\and
		{\rm  Xing Luan}\\
		Southeast University
	} 
	
	\maketitle
	\setlength{\abovedisplayskip}{5pt} 
	\setlength{\belowdisplayskip}{5pt} 
	\begin{abstract}
		The rapid advancement of internet technology has led to a surge in data transmission, making network traffic classification crucial for security and management. However, there are significant deficiencies in its efficiency for handling multi-attribute analysis and its ability to expand model knowledge, making it difficult to adapt to the ever-changing network environment and complex identification requirements. To address this issue, we proposed the SNAKE (Sustainable Network Analysis with Knowledge Exploration) system, which adopts a multi-gated mixture of experts architecture to construct a multi-functional traffic classification model. The system analyzes traffic attributes at different levels through multiple expert sub-models, providing predictions for these attributes via gating and a final Tower network. Additionally, through an intelligent gating configuration, the system enables extremely fast model integration and evolution across various knowledge expansion scenarios. Its excellent compatibility allows it to continuously evolve into a multi-functional large-scale model in the field of traffic analysis. Our experimental results demonstrate that the SNAKE system exhibits remarkable scalability when faced with incremental challenges in diverse traffic classification tasks.  Currently, we have integrated multiple models into the system, enabling it to classify a wide range of attributes, such as encapsulation usage, application types and numerous malicious behaviors. We believe that SNAKE can pioneeringly create a sustainable and multi-functional large-scale model in the field of network traffic analysis after continuous expansion.
	\end{abstract}

	\section{Introduction}
	
With the advent of technologies such as 5G and the Internet of Things, the Internet has undergone rapid and significant expansion. Alongside this growth, new online services, applications, and even cyber threats are continually emerging. According to a recent report by the International Telecommunication Union (ITU), global Internet users reached 4.1 billion in 2019, marking an increase of over $53 \% $ since 2005 and $5.3 \%$ compared to 2018~\cite{report2023}. Ensuring the security and quality of service for Internet users has become a critical concern for major Internet Service Providers (ISPs) and tech companies. Network traffic classification technology is crucial for categorizing Internet traffic based on various criteria, including application type, service, protocol, and potential malicious intent. This technology helps operators deliver customized services, detect cybercriminals, and optimize network resource utilization~\cite{2014Trends}. \par
In the early stages, network traffic was easily identifiable through features like port numbers and identifiers, with the transmitted content being openly accessible and unencrypted. The techniques for classifying network traffic primarily relied on port-based methods ~\cite{2004Transport} alongside DPI(Deep Packet Inspection) which may use certain machine learning algorithms that analyzed the plain-text content within payload~\cite{2010Automatic}. However, with the increasing emphasis on protecting user privacy, encryption protocols, particularly TLS/SSL suites, were developed to secure the data transmitted across networks. This resulted in the widespread use of the HTTPS protocol, an enhanced version of HTTP, which became the preferred choice for network traffic transmission~\cite{10.1007/978-3-662-45670-5_8}. Furthermore, the QUIC protocol, which combines UDP with HTTPS and is especially advantageous for streaming media, has been increasingly adopted. The surge in the variety of applications and services, coupled with the use of port mapping techniques to circumvent firewalls, has rendered the traditional port-based methods for network traffic classification somewhat obsolete. Traffic encryption and dynamic port assignment highlight the growing complexity of managing network traffic, rendering both port-based and payload-based methods ineffective. \par

\begin{figure*}[h]
	\centering
	\includegraphics[width=0.9\textwidth]{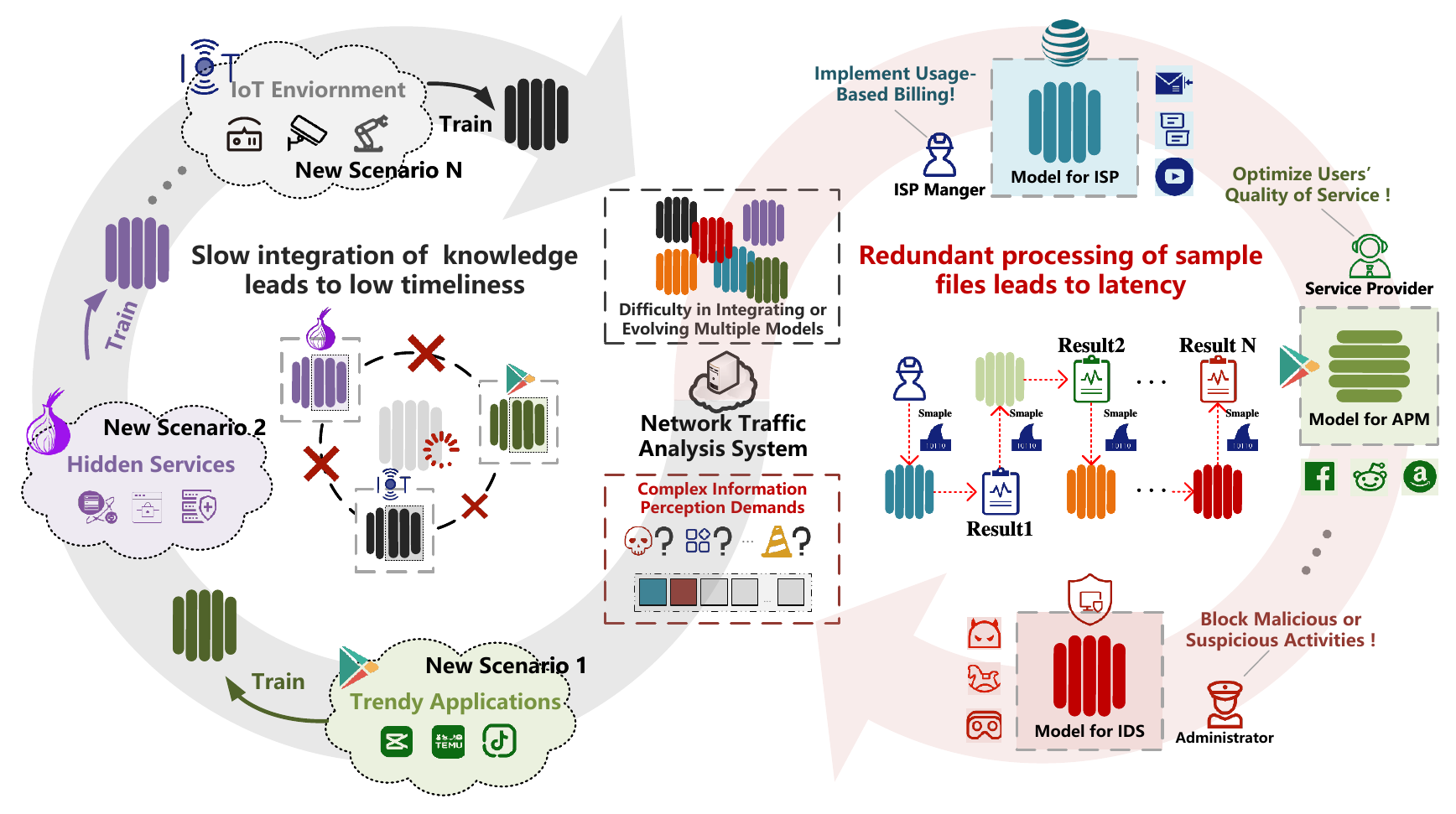} 
	\caption{Challenges in Existing Network Traffic Analysis Systems\\
	\small This diagram illustrates two key deficiencies in current network traffic analysis systems that lead to low timeliness. On the right, the need for diverse classified intelligence requires repetitive processing of traffic samples through multiple models, resulting in inefficiency. On the left, newly emerging sample sets struggle to be quickly integrated or incrementally updated in trained models, further degrading model timeliness.}
	\label{fig1}
\end{figure*}

The underlying traffic data remains interactive and utilizes network media, adhering to specific network protocols. While encryption obscures a significant amount of information, different network services still necessitate formats for segmenting transmitted data, and both parties involved follow established interaction rules. Additionally, various applications develop their own data transmission modules, resulting in differences in the segmentation of traffic data at the application layer (so called application data units~\cite{chen2022a3c}). These characteristics enable traffic classification without the need for decryption. With the ongoing advancements in deep learning, numerous artificial neural network architectures have been developed, many of which (such as GRU, LSTM, and Transformer) are well-suited for processing sequential data. Various features of traffic data can often be represented as discrete sequences, including the lengths of traffic packets ~\cite{FS-net}, the lengths of PDUs/ADUs, and the timing of packet arrivals ~\cite{10.1145/3405672.3409492}, among others. Numerous studies utilizing these models have achieved impressive results in the realm of traffic classification.\par

However, the complexity of traffic classification tasks is heightened by the demand of various subsequent network management functions. For example, in the context of ISP-customized billing ~\cite{ISP} and APM for optimizing the QoS of specific applications ~\cite{khanna2006application}, using website fingerprinting technology for Tor detecting ~\cite{DBLP:conf/ccs/ZhaoD0LL0024}. In systems such as IDS and IPS ~\cite{fu2021realtime}, which are designed to prevent and mitigate malicious attacks, traffic classification must prioritize the detection of harmful behaviors. Additionally, to facilitate content censorship, traffic classification may need to identify flows related to the Tor ~\cite{tor}, VPN ~\cite{vpn}, and specific DNS queries ~\cite{DNS}. Moreover, for particular network scenarios, it may be necessary to develop customized classification models tailored to IoT devices\cite{iot}, smart mobile devices~\cite{mobile}, and Wi-Fi networks~\cite{wifi}. Currently, effective classification models are generally designed for specific tasks and do not offer a unified approach for simultaneously classifying different attribute categories of network traffic.\par

 Figure 1 illustrates two major challenges faced by the current traffic analysis system. Firstly, the network environment is rapidly evolving. Emerging internet services, private protocols, application versions, and corresponding new types of network devices, as well as new families of malware targeting different vulnerabilities, are constantly emerging. This places our traffic analysis technology in a continuously changing and expanding knowledge base, and the models we use must always face the challenge of "concept drift." As shown on the left side of the figure, researchers continuously collect data to train models in various new scenarios. However, these models have different objectives, and their processing logic and feature extraction schemes vary significantly. It is challenging to deeply integrate them with existing models; at most, we can list them in a model pool, which leads to another dilemma. Secondly, network traffic classification technology is often deployed at large-scale network ingress, where traffic is "copied" and transmitted to designated servers for analysis via Network Packet Brokers (NPB) ~\cite{NPB}. This imposes very high requirements on the identification rate of traffic analysis. However, as shown on the right side of the figure, the different services supported by the analysis system have very different focuses on traffic attributes. For example, an Intrusion Detection System (IDS) must swiftly detect attack events; any delay in this information could allow a network attack to succeed, rendering the prior traffic classification ineffective~\cite{jafri2024leo}. Currently, the mixed identification demands can only repeatedly invoke various models in the model pool for classification, resulting in the repeated processing of traffic samples. When faced with high-throughput data, this mode inevitably leads to omissions and even system crashes. \par

To address this issue, we propose the SNAKE(Sustainable Network Analysis with Knowledge Exploration) system, designed for multi-functional traffic classification with robust scalability. The core component of this system is a model architecture that employs a multi-gated mixture of experts (MMoE) approach. This architecture treats traffic classification models for various tasks as distinct expert network sub-networks. Traffic data is uniformly processed by these sub-models and integrated through multiple gating functions, directing different prediction outcomes to specific fully connected layers. The expert model can directly utilize parameters specifically designed for particular tasks, such as application classification, and then fine-tune the fully connected layers to quickly achieve model integration. Unlike regular multi-task learning frameworks, MMoE architecture isolates model parameters for distinct scenarios, minimizing task interference. It also differs from conventional incremental learning approaches by enabling model-level expansion for identical tasks, thereby mitigating the issue of catastrophic forgetting.\par

The contributions of this paper can be summarized as follows:
\begin{itemize}[itemsep=0pt, topsep=0pt]
	\item We have innovatively proposed the SNAKE system, which is an evolutionary framework for traffic analysis that integrates multi-functionality and high scalability. It is currently compatible with five distinct datasets and has accomplished the classification of eight tasks, with performance on each task approaching or even reaching the level of individual models. We believe that it can continuously accommodate various network traffic analysis tasks and generate a universal and efficient large-scale model in this field, significantly enhancing the comprehensiveness, efficiency, and timeliness of the analysis system.	
	\item Specifically, we first adopted a mixture of experts architecture in the traffic classification model. This architecture avoids the direct contribution of parameters across different tasks, allowing the model to effectively accommodate training tasks with significant differences. The separated parameter settings also provide the SNAKE system with strong scalability, enabling it to quickly deploy pre-trained models for new analysis tasks and achieve efficient knowledge exploration.	
	\item Additionally, we drew inspiration from the multi-gate mixture of experts model mechanisms used in other fields, as well as various gating configuration schemes. This allows the SNAKE system to intelligently integrate models under complex scenarios, such as when label relationships are unrelated, parallel, or contain nested relationships, thereby further enhancing its scalability and usability.
\end{itemize}\par
The rest of this paper is organized as follows. Section 2 introduces the background related to encrypted traffic classification and the mixture of experts model. Section 3 presents the various components and operational logic of the SNAKE system. Section 4 describes different task expansion scenarios encountered during actual operations, along with the related mathematical descriptions. In Section 5, we set up various scenarios to evaluate the system's performance. Section 6 discusses its limitations. Finally, Section 7 summarizes the paper. To enable the research community to utilize our tool, we will open source our code upon the paper's acceptance.

	\section{Related Works}
	\subsection{Evolution of Encrypted Traffic Classification Techniques}
	Traditional traffic classification techniques, such as port-based ~\cite{2004Transport} and payload-based ~\cite{2010Automatic} methods, rely on plain-text information or fixed rules in unencrypted traffic. However, dynamic port mapping and the widespread use of encrypted transmission protocols like TLS/SSL have rendered these methods largely ineffective. In response, researchers have turned to machine learning, using labeled traffic datasets and supervised learning models for automated classification. For instance, T. Bujlow and colleagues applied the C5.0 algorithm ~\cite{2012A}, which combines multiple decision trees to enhance accuracy. Similarly, S. Huang and others utilized the k-Nearest Neighbors (kNN) algorithm ~\cite{2009A} to classify internet traffic based on statistical features, bypassing the need for a training phase. Other methods, such as Support Vector Machine (SVM) and Naive Bayes (NB) ~\cite{AZAB2022}, have also been explored. However, traditional machine learning techniques often struggle with over-fitting and lack the capability for fine-grained classification in an open world~\cite{9343185}. This has led to the exploration of deep learning approaches, which offer improved performance and adaptability in complex network scenarios.\par
	With the rapid advancement of deep learning technology, neural network architectures such as residual neural networks have enabled deeper exploration of features related to data packets. For example, Wei et al. ~\cite{7899588}transformed the first 784 bytes of traffic packets from session groupings into image pixels to construct a CNN-based traffic classifier. In 2022, Dai et al.~\cite{9916060} applied the Transformer model architecture, known for its success in natural language processing, to traffic classification with impressive results. More recently,  Wang et al. have established proofs for various mathematical theoretical foundations of deep learning in the field of traffic classification ~\cite{wang2017malware}. Additionally, some deep learning methods for low-quality traffic data have also emerged ~\cite{DBLP:conf/ndss/QingYDCL000024}.\par
	Despite these advancements, the goals of encrypted traffic classification are diverse, including application classification, VPN ~\cite{vpn} and Tor ~\cite{tor} detection, service type identification, and monitoring of malicious traffic such as malware ~\cite{malware}, attacks in IPv6 ~\cite{DBLP:journals/corr/abs-2204-09465}, DDoS attacks ~\cite{ddos}, and even attacks in IoT environments ~\cite{DBLP:conf/uss/Dong0WLZTP0XX23}. However, these models are not well-integrated, posing challenges for multi-objective traffic analysis.\par
	\subsection{Potential Approaches for Scalable Traffic Classification}
	\textbf{Multi-task learning approach:} In the early stages of deep learning, multi-task learning was introduced to tackle various classification tasks within a single model ~\cite{zhang2021survey}. This approach connects multiple fully connected layers, each with its own loss function, to a shared neural network, allowing simultaneous training for different tasks. Some research has applied this method to traffic classification; however, our practical experience suggests limited feasibility for two main reasons:\\  	
	\hspace*{\parindent} 1) Different tasks require distinct feature sets. For instance, identifying unknown protocols demands analyzing local byte entropy changes, while VPN traffic classification relies on detecting packet re-encryption. In contrast, application and service type classification focuses on payload length sequences, and detecting malicious behaviors like DDoS attacks requires attention to statistical traffic features. These diverse features are challenging to integrate into a single model.\\	
	\hspace*{\parindent} 2) Traffic classification attributes can exhibit subordinate and parallel relationships, creating complex task domains ~\cite{parisi2019continual}. For example, traffic may simultaneously belong to a specific protocol, application, and be classified as an attack or associated with VPN/Tor routing. When multi-task learning directly connects these classification targets, task divergence or entanglement can impede effective model training.\par
	\textbf{Large Model Fine-tuning Approach:}Since last year large language models has advanced rapidly, with some of its latest innovations being applied in the field of network ~\cite{zhao2023survey}. However, there has yet to be relevant research that has developed a traffic analysis model capable of achieving multiple classification objectives simultaneously. Current studies typically take one of two approaches: \ding{172}Utilize established open-source large language models, such as Llama and Grok, and develop a model specifically tailored for traffic classification using fine-tuning techniques. \ding{173} Drawing inspiration from the Masked AutoEncoder (MAE) concept found in BERT ~\cite{lin2022bert} and GPT ~\cite{zhang2024trafficgpt}, training a large model head specifically for traffic classification, followed by fine-tuning the terminal layers to perform designated classification tasks ~\cite{dong2024deep}. The first approach, which leverages the parameters of large language models, undoubtedly possesses strong classification capabilities. However, since language models are generally designed as broad conversational tools, most of their parameters are actually irrelevant to traffic classification. This vast number of parameters can result in high computational costs and delays in output during task execution, rendering it less suitable for high-throughput traffic classification scenarios. The second approach appears to be more advantageous, but it primarily aids in training downstream models. Ultimately, these models still need to be executed separately to produce classification results, which does not address the challenges highlighted in the first subsection.\par
	\subsection{Mixture of Experts Architecture}
	The Mixture of Experts (MoE) architecture ~\cite{eigen2013learning} is designed to facilitate the horizontal expansion of models by dynamically routing inputs to different sub-model modules based on actual demands. In neural network models, transforming specific layer outputs into MoE layers allows for the identification of the most suitable expert modules for processing current inputs. By distributing tasks among one or more expert models, the system can efficiently handle a wide array of tasks while maintaining a sparse parameter distribution. This design enables the selective activation of parameters during both training and inference, thereby reducing unnecessary computational power consumption. When applied to traffic analysis, this approach can significantly enhance the speed of result generation.\par
	Additionally, the MoE architecture provides strong horizontal scalability, allowing for the seamless integration of new functional modules into an existing trained model. This capability is especially advantageous in the context of rapidly evolving network traffic scenarios. Advanced variants such as the Multi-gate Mixture of Experts (MMoE) ~\cite{ma2018modeling} further refine this approach by employing multiple gating networks to allocate inputs to the most relevant experts, enhancing model performance in multi-task learning environments. Other techniques, such as Hierarchical MoE ~\cite{li2024hierarchical} and Switch Transformers ~\cite{fedus2022switch}, provide additional flexibility and efficiency in routing and processing, demonstrating the versatility of MoE architectures in adapting to complex and dynamic classification tasks.
	
	\vspace{-8pt}\section{System Design of SNAKE}
	\begin{figure*}[h]
		\centering
		\includegraphics[width=0.8\textwidth]{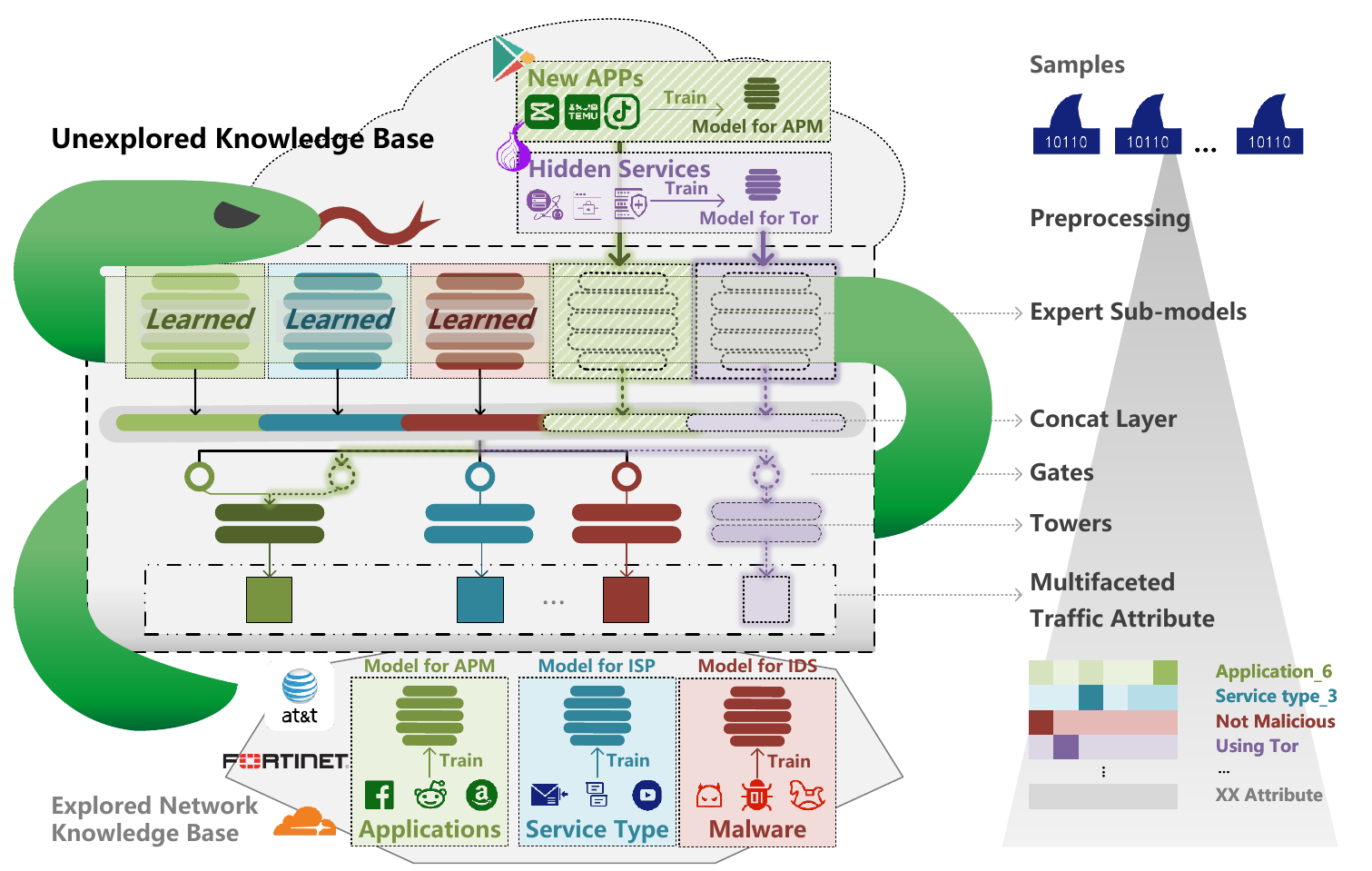} 
		\caption{Overview of the SNAKE System Structure \\
			\small The SNAKE system is designed to efficiently aggregate various network traffic classification models. It comprises expert sub-models from pre-trained models in different domains, a concatenation layer that combines their inputs, gates that control the flow of information, and tower layers for specific tasks. This system can continuously integrate new task models, enabling rapid classification of multiple network traffic attributes.}
		\label{fig2}
	\end{figure*}
	\subsection{Threat Model}
		As discussed, the current landscape of traffic classification tasks is exceedingly diverse, with many tasks playing a crucial role in network security and management. While the deployment of classification systems aims to provide a comprehensive understanding of data flows, there is a lack of a cohesive framework to integrate techniques from various domains. A single model often struggles to meet the complex demands of different scenarios. For instance, intrusion detection systems must address malicious data from various environments, including mobile devices, IoT, even in devices using microphone~\cite{20243917120586}, as well as diverse attack patterns such as DDoS~\cite{DBLP:conf/sp/LiWDL24} and malicious DoH .\par
	
	Moreover, common user traffic can experience concept drift due to changes in internet protocols and emerging services, impacting the overall classification effectiveness of the model. The significant disparities among different tasks further impede multi-task learning architectures designed to accommodate multiple objectives. This challenge stems from the direct sharing of underlying model parameters ~\cite{eigen2013learning}, which restricts their ability to meet the varied requirements of distinct tasks.\par
	
	The objective of the SNAKE system is to facilitate rapid model fusion and multi-target identification in the complex domain of network traffic analysis. Our final implementation enables the architecture trained by the SNAKE system to classify various application types, service categories, VPN and Tor usage, as well as identify four major categories of Android malware (including 42 specific attack tools), three types of DoH attacks, and ten types of Trojans. Achieving these functionalities requires the SNAKE model to exhibit strong scalability and specific handling mechanisms for compatibility across different task patterns.
	\vspace{-8pt}\subsection{Overview}
	To construct a highly scalable traffic classification system compatible with multiple identification targets, SNAKE draws on the MMoE architecture to build its classification model. At the model hierarchy level, SNAKE consists of four important components. First, all traffic classification tasks share a unified data preprocessing system for feature extraction. To accommodate different tasks, we have referenced the work of Distiller ~\cite{distiller}, extracting information from traffic headers and payloads to form a common input for our model.\par
	The second component is the combination of expert networks, which are tailored for different traffic classification tasks. The SNAKE system can set up the same network structure to directly read pre-trained model parameters. Certainly, the model architecture also supports the option to train a specific classification task from the ground up; however, due to the complexity of the overall network structure, we do not recommend adopting this approach.\par
	The third component comprises an intelligent multi-gating structure featuring configurable modes, which is employed to regulate the outputs of multiple expert networks tailored for specific tasks. We have developed three distinct gating schemes corresponding to three different task fusion scenarios, with detailed descriptions to be provided in subsequent sections.\par
	Finally, the architecture includes several classification tower networks designed for various models, which typically comprises fully connected networks. Following the model fusion process, only a limited number of fine-tuning iterations will be conducted on this segment of the model parameters. The overall architecture of the SNAKE system is depicted in Figure 2.
	
	\vspace{-8pt}\subsection{Preprocessing}
	Our method adopts a traffic preprocessing identifying individual flows as the fundamental classification unit, uniquely defined by their five-tuple. The input data provided to the classifier comprises two types: (a) the first Nb bytes of the transport-layer payload (PAY) of the Traffic Classification object ~\cite{wang2020id}; and (b) informative protocol header fields (HDR) from the first Np packets ~\cite{lopez2017network}.
	In the first case, the input is represented in binary format, organized in a byte-wise manner, and normalized within the range of (0, 1). The second type of input data includes: (i) the number of bytes in the transport-layer payload, (ii) the TCP window size (which is set to zero for UDP packets), (iii) inter-arrival time, and (iv) packet direction $ \in {0,1} $ derived from the first  $N_p$ packets.\par
	It is important to note that, in both scenarios, longer instances are truncated while shorter ones are padded with zeros to conform to the specified lengths of bytes ($N_b$) or packets ($N_p$). This specific input data selection is driven by the need to mitigate biased inputs—a common pitfall identified in related works—which may arise from factors such as PCAP metadata, data-link layer information, and certain transport-layer header fields (e.g., source and destination ports). Such biases can lead to inflated performance metrics and hinder generalization capabilities.\par
	This processing approach enables us to accommodate both flow-level and packet-level perspectives, thereby supporting a range of tasks with varying emphases. For instance, while tasks related to DDoS attacks or port scanning require a flow-centric view, application classification tasks necessitate a detailed analysis of the payload data.\par
	\vspace{-4pt}\subsection{Expert Sub-models}
	The expert classification networks served as the primary entity for executing a variety of tasks. In this paper, we adopt a fine-tuning approach for the integration of expert sub-models, wherein we first independently train usable models on their respective labeled datasets. Subsequently, the SNAKE system retrieves and freezes the parameters of the core components from these models, relying solely on the gating mechanisms and tower networks in the latter stages for fine-tuning. It is acknowledged that there exists a plethora of model architectures tailored for different classification tasks, with extensive research conducted in the field. However, the diverse input formats of these models are evidently detrimental to the model fusion process. In light of the data preprocessing methods discussed in the previous section, this paper proposes a unified architecture for the expert sub-models.\par
	In the original Distiller work, two modalities of input were analyzed using GRU and convolutional neural networks, which were effective for capturing temporal dependencies and texture features of payload data, respectively. However, with the advent of the transformer architecture, we have identified that this framework can effectively accommodate both modalities, thereby simplifying the model structure and avoiding the complexities arising from the use of heterogeneous architectures. The transformer architecture employs a multi-head attention mechanism composed of queries (Q), keys (K), and values (V), enabling it to extract sequential dependencies across varying spans, thus fulfilling the analytical requirements for both flow-level and packet-level modalities. In this paper, we uniformly adopt an expert model structure that flattens and concatenates the data from both modalities as input, establishing two attention heads for feature extraction. The specific model structure parameters and hyper-parameters utilized for training each traffic classification task are presented in Table 1.
	
	\begin{table}[ht]
		\centering
		\caption{\small Architecture and Hyper-parameters of Expert Sub-models}
		\small 
		\begin{tabular}{@{}ll@{}}
			\toprule
			\textbf{Parameter} & \textbf{Value} \\ \midrule
			\; \ Transformer Encoder & 912dim, 2 heads \\ 
			\; \ Linear Layers & [912, 256, $N_{\text{target}}$] \\ 
			\multicolumn{1}{l}{Optimizer, Learning Rate, Activation} & Adam, 1e-3, ReLU \\ 
			\multicolumn{1}{l}{Dropout Rate, Batch Size, Epochs} & 0.2, 32, 50 \\ 
			\; \ Loss Function & Cross-Entropy \\ 
			\bottomrule
		\end{tabular}
		\label{tab:nn_hyperparameters}
	\end{table}
	
	During the model fusion process within the SNAKE system, only the parameters of the core transformer model components are read, excluding the final fully connected layers. We would like to emphasize that there may exist more optimal data processing or model configuration strategies for each specific task. However, this paper primarily focuses on evaluating the model fusion and expansion capabilities of SNAKE; hence, we have selected a relatively generic architecture for this purpose. The system itself supports customization of the structure, parameters, and hyper-parameters for each expert network.\par

	\vspace{-4pt}\subsection{Gates and Towers}
	In this paper, we adopt a supervised learning paradigm for traffic classification. In the field of network traffic research, labeled sample data suitable for various tasks, such as malicious attacks and traffic application types, can be collected using constructed scripts and some automated tools. Thus, dataset can be described as  $D_k(x_i,y_{i,k})$. The subscript $k$ denotes the classification task for the $k_{th}$ attribute, where $x_i$ represents the $i_{th}$ traffic sample and $y_{i,k}$ is the corresponding label. We build the SNAKE system to provide multiple attribute categories of traffic at once, and this system employs a MMoE architecture, establishing an independent gating layer for each task. Therefore, for the known classification task of $k$ attributes, we need to establish $k$ Gates and corresponding Tower classification networks for the dataset. We assume that $n$ expert sub-models $f_j(x)$ have been used previously, and we need to define a concatenation operation to represent it as follows:
	$$ X_i = concat(f_1(x_i),f_2(x_i), \dots ,f_n(x_i)) = \left[
	\begin{array}{c}
		f_1(x_i) \\
		f_2(x_i) \\
		\vdots \\
		f_n(x_i)
	\end{array} \right]$$
	
	After the $k_{th}$ gate $g_k$ receives the overall output $X$ , it needs to be integrated and input into the corresponding Tower $h_k$. Specific operation can be represented as follows:
	$$ y_k = h_k(g_k(X_i)) $$	\par
	In the conventional MMoE architecture, the gate is a learnable linear transformations of the input with a softmax layer. However, in the field of traffic classification, whether the outputs of different expert models are mixed depends on the correlation between the corresponding task attributes. Therefore, the gating we use is merely a vector of the same dimension as the traditional MMoE method, with a length equal to the number of expert models $n$, like $g_i(\delta_{1i},\delta_{2i}, \dots,\delta_{ni})$. For the specific value settings, we have the following three modes:\par
	\vspace{0.2cm}
	\textbf{Default Mode:} In this mode, if there is only one expert model connected to the gate, the task and the expert are in a one-to-one correspondence. Consequently, the gating mechanism is configured to output only the result from the corresponding expert, as expressed below:
	$$ 
	g_i(\delta_{1i},\delta_{2i}, \dots,\delta_{ni}),\qquad where \; \ \delta_{ji} =
	\begin{cases}
		1 & \text{if } j = i \\
		0 & \text{if } j \neq i
	\end{cases}
	 $$
	\par
	\textbf{Top-K Mode:} In this mode, multiple expert models are connected to the gate, and these experts share the same classification granularity. For instance, Expert 1 may classify $app_1$, $app_2$, ..., $app_n$, while Expert 2 classifies $app_{n+1}$, ..., $app_m$. Here, the task and experts are in a one-to-many relationship, and the gating mechanism is designed to equally consider the results from those experts (which could be represent by a set $S_j \subseteq \{1, 2, \ldots, n\}$).
	The setting of the gating vector in the top-k mode is as follows:
	$$g_i(\delta_{1i},\delta_{2i}, \dots,\delta_{ni}),\qquad where \; \ \delta_{ji} =
	\begin{cases}
		1/n & \text{if } j \in S \\
		0 & \text{if } j \notin S
	\end{cases}$$
	\par
	
	\begin{algorithm}[ht]
		\caption{Multi-Attribute Traffic Classification Process Using the SNAKE System}
		\label{alg:traffic_classification}
		\begin{algorithmic}[1]
			\STATE \textbf{Notation:} \( X \): Traffic sample, \( Y_j \): Attribute value for task \( j \), \( N \): Number of experts, \( K \): Number of classification tasks, \( HDR \): packet headers feature vector, \( PAY \): Payload feature vector, \( V \): Preprocessed data vector, \( V_i \): Output from expert model \( E_i \), \( \hat{V} \): Concatenated outputs, \( g_j \): Gating value for task \( j \), \( G_j \): Gated output for task \( j \), \( T_j \): Tower network for task \( j \), \( C_j \): Confidence vector for task \( j \)
			
			\STATE \textbf{Input:} Traffic sample \( X \)
			\STATE \textbf{Output:} Attribute values \( Y_1, Y_2, \ldots, Y_K \)
			\STATE 
			\COMMENT{Step 1: Preprocess the traffic data}
			\STATE Extract \( HDR \) and \( PAY \) from \( X \)
			\STATE Normalize and encode the extracted data
			\STATE 
			\COMMENT{Step 2: Mixture of Experts Model}
			\STATE Concatenate preprocessed data into vector \( V \)
			\FOR{each \( E_i \) in \( N \)}
			\STATE Pass \( V \) through \( E_i \)
			\STATE Store output \( V_i \)
			\ENDFOR
			\STATE 
			\COMMENT{Step 3: Gating Mechanism}
			\STATE Concatenate outputs: \( \hat{V} = \text{concat}(V_1, V_2, \ldots, V_N) \)
			\FOR{each \( T_j \) for \( j = 1, 2, \ldots, K \)}
			\STATE Compute value \( g_j \) for task \( j \)
			\STATE Apply \( g_j \) to \( \hat{V} \) to produce \( G_j \)
			\ENDFOR
			\STATE 
			\COMMENT{Step 4: Tower Networks}
			\FOR{each \( T_j \) for \( j = 1, 2, \ldots, K \)}
			\STATE Pass  \( G_j \) to \( T_j \)
			\STATE Generate \( C_j \) for task \( j \)
			\ENDFOR
			\STATE 
			\COMMENT{Step 5: Extract Attribute Values}
			\FOR{each \( C_j \)}
			\STATE \( Y_j = \text{argmax}(C_j) \) \quad \text{(Attribute value for task \( j \))}
			\ENDFOR
			\STATE 
			\RETURN Attribute values \( Y_1, Y_2, \ldots, Y_K \)
		\end{algorithmic}
	\end{algorithm}
	
	\textbf{Trainable Mode:} In this mode, multiple expert models are connected to the gating layer, and these experts operate at different classification granularities. For example, Expert Model 1 may judge whether this traffic is malicious, while Expert Model 2 focuses on specific tools. In this case, there exists a many-to-many relationship between the tasks and the experts. If the gating layer were to equally consider the outputs of both models, it could interfere with their respective tasks. Therefore, these gating layers should be set to an updatable state, allowing for timely updates during fine-tuning based on the gradients returned by the loss function. Specifically, the gating vector can be updated using the following formula: 
	$$g_i(\delta_{1i},\delta_{2i}, \dots,\delta_{ni}),\qquad where \; \ \delta_{ji} =
	\begin{cases}
		\omega_j & \text{if } j \in S \\
		0 & \text{if } j \notin S
	\end{cases}$$
	where $\omega$ is the trainable weight calculated through a linear layer and softmax function, ensuring that the gating layer can flexibly adapt to the outputs of different expert models, thereby optimizing overall classification performance.\par

	Each gating layer connects to the final tower network for the task, which is responsible for producing the ultimate output. For this purpose, we uniformly configure a fully connected layer network. Additionally, to address potential over-fitting during fine-tuning, we have incorporated dropout layers. Specifically, tower networks are configured as a two-layer fully connected network, utilizing a dropout rate of 0.2, a learning rate of 0.001, a batch size of 128, and the ReLU activation function. \par	
	It is important to note that the architecture, parameters, and hyper-parameters of each tower can also be customized. Through experimentation, we have found that adding more fully connected layers does not significantly impact model performance; thus, we have opted for a simplified configuration in this context.\par
	The pseudo-code for the system's operation is presented in Algorithm 1.

	\vspace{-12pt}\section{Mathematical Description and Convergence Proof of Incremental Knowledge Scenarios}
	
	In the previous section, we introduced the three modes set by Gates, which correspond to three different task integration scenarios in the field of network traffic classification. These three scenarios are somewhat similar to the task expansion learning scenarios mentioned in the literature ~\cite{van2022three}. However, our assumption here is that we have a clear understanding of the label attributes of the traffic samples and their relationships with other attributes.\par
	
	We define that for a certain traffic sample, there exists its corresponding task domain and label domain. We have selected several examples as shown in the Figure 3.
	
	\begin{figure}[h]
		\centering
		\includegraphics[width=0.45\textwidth]{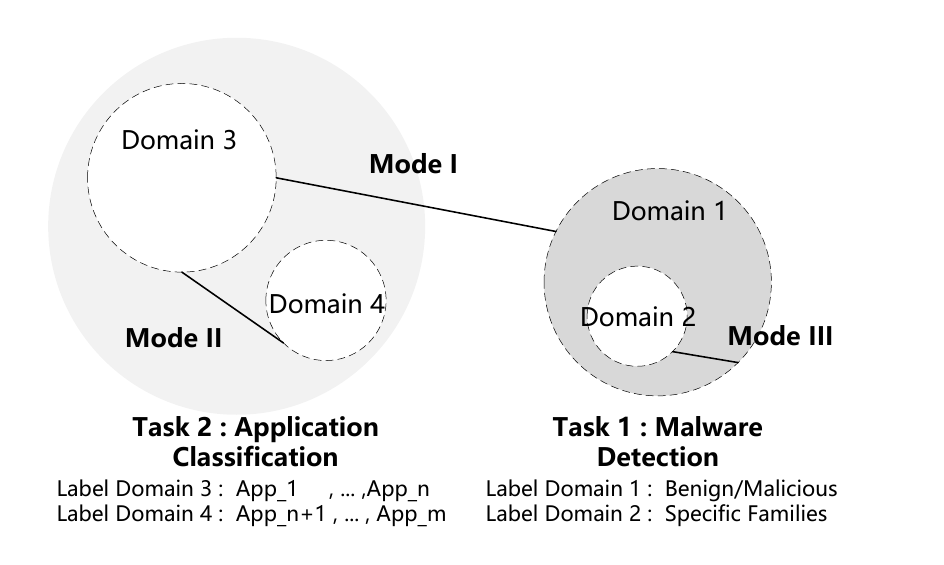} 
		\caption{Examples of Task and Label Domains \\ 
			\small This figure illustrates the examples of corresponding task domains and label domains, along with the relationships between them.}
		\label{fig3}
	\end{figure}
	
	We define $ \mathcal{X} $ as the sample input space, $\mathcal{Y}$ as the corresponding label space for classification, and $\mathcal{T}$ as the function that implements the mapping between them. As shown in the figure, we have the following three situations when handling task fusion.\\
	\textbf{Task Attribute-Independent Scenarios:} As shown in Mode I in the figure, tasks of distinguishing whether the traffic is malicious and classifying the traffic based on the application set in label domain 3 is mutually independent.In this case, our task fusion objective can be represented as follows:
	\[
	\begin{array}{rl}
		\text{Given} &\mathcal{X} \times \mathcal{T}_1 \rightarrow \mathcal{Y}_1, \quad \mathcal{X} \times \mathcal{T}_2 \rightarrow \mathcal{Y}_2  \\
		\text{Search} & \mathcal{F} \quad \text{s.t.} \quad \mathcal{F}(\mathcal{T}_1, \mathcal{T}_2): \mathcal{X} \xrightarrow{} (\mathcal{Y}_1, \mathcal{Y}_2)
	\end{array}
	\] 
	
	\vspace{-4pt} \textbf{Category Expansion Scenarios:} As shown in Mode II, the classification tasks from $app_1$ to $app_n$ and from $app_{n+1}$ to $app_m$ belong to the category expansion within the same task, which can be expressed as follows:
	\[
	\begin{array}{rl}
		\text{Given} &\mathcal{X}_{L_1} \times \mathcal{T}_1 \rightarrow \mathcal{Y}_{L_1}, \quad \mathcal{X}_{L_2} \times \mathcal{T}_2 \rightarrow \mathcal{Y}_{L_2}  \\
		\text{Search} & \mathcal{F} \quad \text{s.t.} \quad \mathcal{F}(\mathcal{T}_1, \mathcal{T}_2): \mathcal{X}_{L_1 \cup L_2} \xrightarrow{} \mathcal{Y}_{L_1 \cup L_2}
	\end{array}
	\]
	
	\vspace{-4pt} \textbf{Category Refinement Scenarios:} As shown in Mode III, the tasks of determining whether a classified traffic is malicious and identifying the specific malicious family category of the flow represent a relationship where the former task is refined. This can be expressed as follows:   
	\[
	\begin{array}{rl}
		\text{Given} &\mathcal{X} \times \mathcal{T}_1 \rightarrow \mathcal{Y}, \quad \mathcal{X} \times \mathcal{T}_2 \rightarrow \mathcal{Y}_1 \quad (\mathcal{Y}_1 \in \mathcal{Y}) \\
		\text{Search} & \mathcal{F} \quad \text{s.t.} \quad \mathcal{F}(\mathcal{T}_1, \mathcal{T}_2): \mathcal{X} \xrightarrow{} \mathcal{Y}\xrightarrow{} \mathcal{Y}_1
	\end{array}
	\]
	
		\begin{algorithm}
		\caption{SNAKE System Training Process for Model Task Expansion}
		\begin{algorithmic}[1]
			\STATE \textbf{Definitions:} \(T_1\): Model for Task1; \(T_2\): Model for Task2; \(X_1, Y_1\): Dataset and labels for Task1; \(X_2, Y_2\): Dataset and labels for Task2; \(g_1, g_2\): Gates for each task; \(\mathcal{T}\): Tower Network; \(M\): New MMoE Model
			\STATE \textbf{Input:} \(T_1, T_2\); \(X_1, X_2\); \(Y_1, Y_2\)
			\STATE \textbf{Output:} \(M\)
			\STATE \textbf{Initialize components of model \(M\):}
			\STATE SET \(M.experts\) = \({T_1,T_2}\)
			\STATE Analyze \(Y_1\) and \(Y_2\) to determine Mode
			\STATE \textbf{Switch mode:}
			\STATE \quad \textbf{case Mode I:}
			\STATE  Configure \(g_1, g_2\) as gates(default mode); \\
			\(M.gate \gets (g_1, g_2)\); Configure \((\mathcal{T}_1, \mathcal{T}_2)\); \(M.tower \gets \mathcal{T}\) 	
			\STATE \quad \textbf{case Mode II:}
			\STATE  Configure \(g_1\) as gate(top-k mode);\\
			\(M.gate \gets g_1\); Configure \(\mathcal{T}_1\); \(M.tower \gets \mathcal{T}\) 	
			\STATE \quad \textbf{case Mode III:}
			\STATE  Configure \(g_1, g_2\) as gates(trainable mode);\\
			\(M.gate \gets (g_1, g_2)\); Configure \((\mathcal{T}_1, \mathcal{T}_2)\); \(M.tower \gets \mathcal{T}\) 
			\STATE Lock \(M.experts\)
			\STATE Fine-tune \(M.tower\) with datasets \(X_1, Y_1\) and \(X_2, Y_2\)	
			\STATE Update \(M.tower\) after several epochs
		\end{algorithmic}			
	\end{algorithm}
	
	After defining and describing the above scenario, the task flow of the SNAKE system in executing specific scenarios is represented as Algorithm 2. Of course, in practical scenarios, the categories in Mode II may overlap; in this case, the output dimension of the Tower network used to set the base should be the same as the total number of sample types.\par

	\vspace{4pt}
	
	\textbf{Convergence analysis:} The SNAKE system employs a mixed expert model architecture with separate parameter updates. One reason for this is to avoid the difficulty of model convergence caused by overlapping loss functions from multiple objectives updating shared model parameters. In the architecture we adopted, each cross-entropy function only updates the parameters of its connected Tower network, ensuring that the training updates for different tasks are independent of each other.\par 
	Assuming we have $N$ expert networks, each expert $n$ handles a different classification task or objective. The output of each expert is $\hat{y}_n $ , and these outputs will be concatenated into a single input, which is then passed to gates layer, represented as: $ g = [\hat{y}_1, \hat{y}_2, \ldots, \hat{y}_N]$. For each classification target $k$, we define a cross-entropy loss function $L_k$ that depends on the output from the gating network and the true label $y_k$ :
    $$L_k(y_k, \hat{y}) = -\sum_{i} y_{k,i} \log(\hat{y}_i)$$	
	In this case, the overall loss function can be expressed as a weighted sum of the losses from all classification targets:
	$$
	L_{\text{total}} = \sum_{k=1}^{K} \alpha_k L_k(y_k, \hat{y}_{\text{final}})
	$$
	Here, $\alpha_k$ is the weight for each classification target, which can be adjusted by gates. Because we have the configuration scheme for the Gates as described in Section 3, it ensures that $\alpha_k$ is a non-negative number. This allows us to conclude that the overall loss function can be regarded as a linear combination of several cross-entropy losses, which ensures that it can maintain the Lipschitz continuity and strong convexity of the cross-entropy function when processing softmax outputs. This guarantees that we can present the following lemma, where we set our model output as $\omega$.\\
	\textbf{Lemma 1} (Lipschitz-Continuity) : $L_{total}(\omega)$ (represent as $L_o$) is continuously differentiable and its corresponding $\nabla L_t$ is Lipschitz continuous with constant $c > 0$, that is to say $\forall \omega_1, \omega_2 \in \mathbb{R}^d$ s.t. :
	\begin{equation*}
		\resizebox{0.6\hsize}{!}{$
			\parallel \nabla L_o(\omega_1)-\nabla L_o(\omega_2) \parallel _2 \leq c \parallel \omega_1 - \omega_2 \parallel_2 $}
	\end{equation*} 
	\textbf{Lemma 2} (Strong Convexity) : $L_{o}(\omega)$ is strongly convex, that is to say $\forall \omega_1, \omega_2 \in \mathbb{R}^d$, $ \exists c > 0$ s.t. :
	\begin{equation*}
		\resizebox{0.8\hsize}{!}{$
			L_o(\omega_2)- L_o(\omega_1) \leq \nabla L_o(\omega)^\top (\omega_1 - \omega_2) + \frac{1}{2} \xi \parallel \omega_2 - \omega_1 \parallel^2_2 $}
	\end{equation*} 
	This lemma elucidates that $L_o(\omega)$ is characterized by a singular, unique global minimum. Here,  we designate the point as $L_o(\omega^*)$, where $\omega^*\in \mathbb{R}^d$. Based on these two lemmas, we can ultimately infer one conclusion. When setting  the learning rate reasonably, the overall loss function will eventually approximate its minimum value. After $T$ rounds of training, setting learning rate $\alpha \leq \frac{1}{c}$,  the convergence upper bound of $L_o(\omega)$ can be formulated as follows:
	\begin{equation*}
		\resizebox{0.45\hsize}{!}{$ L_o(\omega^T)-L_o(\omega^*) \leq \frac{1}{Tz\alpha(1-\frac{c\alpha}{2})}
			$}
	\end{equation*} 
	This indicates that as the number of training iterations increases, $L_o(\omega^T)$ will gradually converge to the global minimum $L_o(\omega^*)$. For the sake of the article's coherence, the complete proof can be found in \textit{Appendix Section A} . \par
	
	\section{System Evaluation}
	
	We established eight distinct traffic classification tasks, encompassing five public datasets with a total data volume of 172.29 GB, to validate the effectiveness of the SNAKE system. The experiments include a comparative analysis of the structural configurations of the expert models, an assessment of the model fusion effectiveness across three task expansion scenarios, a comparison of model hyper-parameter settings, and, finally, an evaluation of integrating numerous tasks scenario.
	
	\vspace{-26pt}\subsection{Experimental Setup}
	In this paper, aside from the corresponding comparative experiments, the default expert network structure and parameters, tower model parameters, dropout settings, and other hyper-parameter configurations are consistent with those outlined in Section 3. Some brief information about the datasets used in our experiments can be found in Table 3. Due to the use of multiple datasets, a detailed introduction is provided in \textit{Appendix Section B} for the sake of narrative clarity.
	
	\begin{table}[ht] 
		\centering
		\caption{Overview of the Datasets Used}
		\scalebox{0.75}{ 
			\begin{tabular}{@{}lccr@{}}
				\toprule
				Dataset Name & Classes & Dataset Size & Task Objective \\ \midrule
				ISCXVPN2016~\cite{VPN2016}    & 2/6/15    & 8.58G & VPN/Service/Application  \\
				ISCXTor2016~\cite{Tor2016}    & 2      & 21.6G  & Tor usage \\
				CIC-DoHBrw-2020~\cite{DoH}     & 2      & 90.2G  & Malicious DNS \\
				USTC-TFC2016~\cite{7899588}     & 11      & 3.71G  & Malware Detection  \\
				IPTAS-Tbps~\cite{chen2022a3c}    & 7     & 10.7G & Application Classification  \\ \bottomrule
			\end{tabular}
		}
		\label{tab:datasets}
	\end{table}
	\par
	
	During the model training, we performed a uniform split of the dataset, allocating $75\%$ to the training set, $10\%$ to the validation set, and $20\%$ to the test set. Whether it is the training of the expert model or the fine-tuning and integration operations conducted by the SNAKE system, the training samples are consistently drawn from the training set. Therefore, the experimental results will not be misjudged due to any overlap in the datasets. We conducted experiments with ten random data splits, so all the following experiments present the results of these ten repeated trials.\par
	\begin{table*}[h]
	\captionof{table}{Performance of the proposed expert sub-model structures across different tasks}
	\small
	\centering
	\scalebox{0.73}{
		\begin{tabular}{lccccccccc}
			\toprule
			\multirow{2}{*}{Algorithm} & \multicolumn{3}{c}{Encapsulation} & \multicolumn{3}{c}{Traffic Type} & \multicolumn{3}{c}{Application} \\
			\cmidrule(lr){2-4} \cmidrule(lr){5-7} \cmidrule(lr){8-10}
			& Accuracy (\%) & Precision (\%) & F1 Score (\%) & Accuracy (\%) & Precision (\%) & F1 Score (\%) & Accuracy (\%) & Precision (\%) & F1 Score (\%) \\
			\midrule
			SNAKE Expert & \textbf{98.66($\pm$ 2.22)} & \textbf{98.67 ($\pm$ 2.18)} & \textbf{98.67 ($\pm$ 2.22)} & \textbf{80.25 ($\pm$ 1.19)} & \textbf{80.01 ($\pm$ 3.90)} & \textbf{ 79.65 ($\pm$ 1.05)} & \textbf{77.28($\pm$ 0.87)} & \textbf{77.75($\pm$ 2.78)} & \textbf{75.84 ($\pm$ 1.38)} \\
			CNN-RNN-2a~\cite{algorithm1} & \( 65.63 \sim 90.00 \) & \( 74.11 \sim 90.07 \) & \( 58.22 \sim 90.02 \) & \( 69.46 \sim 75.62 \) & \( 72.69 \sim 78.94 \) & \( 67.66 \sim 74.31 \) & \( 52.05 \sim 67.01 \) & \( 55.74 \sim 69.86 \) & \( 47.44 \sim 63.93 \) \\
			MTLCNN~\cite{algorithm3} & \( 82.60 (\pm 1.71) \) & \( 83.27 (\pm 1.74) \) & \( 82.65 (\pm 2.05) \) & \( 73.78 (\pm 2.68) \) & \( 77.10 (\pm 3.66) \) & \( 72.17 (\pm 2.84) \) & \( 63.46 (\pm 7.89) \) & \( 65.56 (\pm 9.59) \) & \( 59.55 (\pm 10.81) \) \\
			MT-DNN-FL~\cite{algorithm4} & \( 82.63 (\pm 8.91) \) & \( 83.03 (\pm 7.26) \) & \( 81.82 (\pm 9.86)\ \) & \( 74.05 (\pm 2.68) \) & \( 77.26 (\pm 3.66) \) & \( 72.35 (\pm 2.84) \) & \( 64.56 (\pm 3.68) \) & \( 66.95 (\pm 4.17) \) & \( 59.94 (\pm 10.43) \) \\
			multi-output DNN~\cite{algorithm5} & \( 97.98 (\pm 0.41) \) & \( 97.99 (\pm 0.37) \) & \( 97.99 (\pm 0.41) \) & \( 77.78 (\pm 2.23) \) & \( 78.55 (\pm 2.83) \) & \( 76.47 (\pm 2.60) \) & \( 75.35 (\pm 1.88) \) & \( 76.45 (\pm 3.04) \) & \( 74.38 (\pm 1.64) \) \\
			\bottomrule
	\end{tabular}}
	\caption*{\scriptsize \hfill * All results are presented as mean $\pm$ range, except for CNN-RNN-2a}
\end{table*}

\begin{figure*}[htbp]
	\centering
	\begin{subfigure}[b]{0.32\textwidth}
		\centering
		\includegraphics[width=\textwidth]{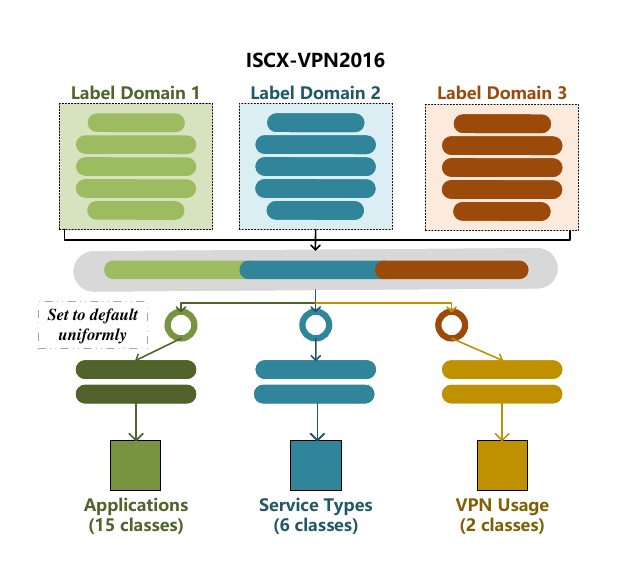} 
		\caption{Task Attribute-Independent Scenarios}
		\label{fig:sub_a}
	\end{subfigure}
	\hfill
	\begin{subfigure}[b]{0.32\textwidth}
		\centering
		\includegraphics[width=0.58\textwidth]{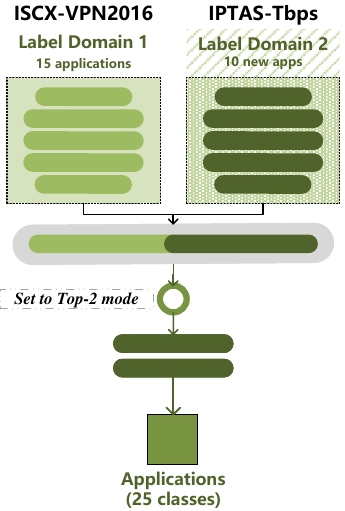} 
		\caption{Category Expansion Scenarios}
		\label{fig:sub_b}
	\end{subfigure}
	\hfill
	\begin{subfigure}[b]{0.32\textwidth}
		\centering
		\includegraphics[width=0.72\textwidth]{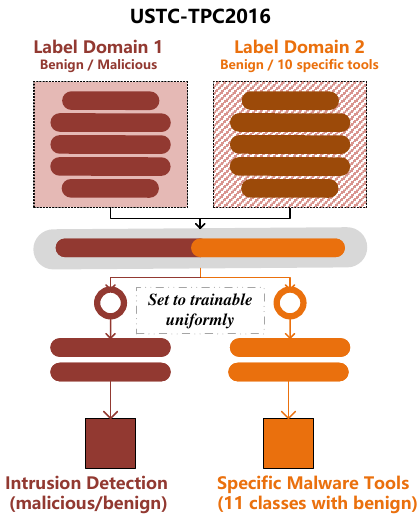} 
		\caption{Category Refinement Scenarios}
		\label{fig:sub_c}
	\end{subfigure}
	
	\caption{Model Extension Performance in Three Incremental Knowledge Scenarios}
	\label{fig:model_extension}
\end{figure*}
		
	In our experiments, we utilized deep learning models implemented with PyTorch 2.0.0 and CUDA 11.7. Data pre-processing and post-processing were primarily conducted using the NumPy and Scapy libraries. For graphical data representation, we employed Matplotlib and MATLAB. All experiments were performed on a PC with the following hardware specifications: a 13th Gen Intel® Core™ i7-13700KF processor running at 3.40 GHz, 32 GB of RAM, and an NVIDIA GeForce RTX 4080 GPU. The operating system used was Windows 11.\par

	\subsection{Experiments on Classification Effects of Different Expert Models}
	In this subsection, we aimed to compare the recognition performance of expert networks using different model architectures. Here, we compare five traffic classification models that utilize data processing schemes similar to ours. We evaluated the detection results of these models on the ISCX-VPN-2016 dataset for three tasks: encapsulation, service type, and application, as shown in Table 3.\par
	
	In this experiment, all models were set with a batch size of 16 and trained for 50 epochs. The dataset was randomly split with 75$\%$ for the training set, 10$\%$ for the validation set, and 15$\%$ for the test set, and this process was run ten times. The table shows the accuracy, precision, and F1-score values of five model architectures. The CNN-RNN-2a model exhibited significant oscillations, possibly because the small batch size and number of epochs were not well-suited for training RNN models. Therefore, its range of variation is presented in the table, while the results of the other models are shown as average results and range performance. It is evident that the expert model architecture adopted in this paper performed the best in this three tasks, outperforming the other optimal results by approximately 0.5$\%$ to 2.5$\%$ on average. According to the experimental results, the transformer method we employed demonstrates significantly superior overall performance across the three tasks. Therefore, we used this model architecture for all expert networks in the classification tasks.
	
	\vspace{-8pt}\subsection{Model Extension Performance in Three Incremental Knowledge Scenarios}
	To explore the optimal gating configuration pattern and to lay the groundwork for the SNAKE system to integrate complex traffic classification task sets, we set up experiments in three traffic classification task fusion scenarios to test the effectiveness of different gating configurations. The schematic diagram of the scenarios is shown in Figure 4.

	\vspace{-8pt}\subsubsection{Task Attribute-Independent Scenarios}
	In this scenario, we utilized the three classification tasks from ISCX-VPN-2016 to construct a task attribute-independent model fusion scenario. As shown in Figure 4 (a), the SNAKE system established three gates and corresponding Tower networks for the classification of three traffic attributes in the ISCX-VPN-2016 dataset. All gates were set to default mode which means the base Tower network only considered the outputs of corresponding domain expert sub-models. 
	
	\begin{figure}[h]
	\centering
	
	\begin{minipage}{0.235\textwidth} 
		\centering
		\includegraphics[width=\textwidth]{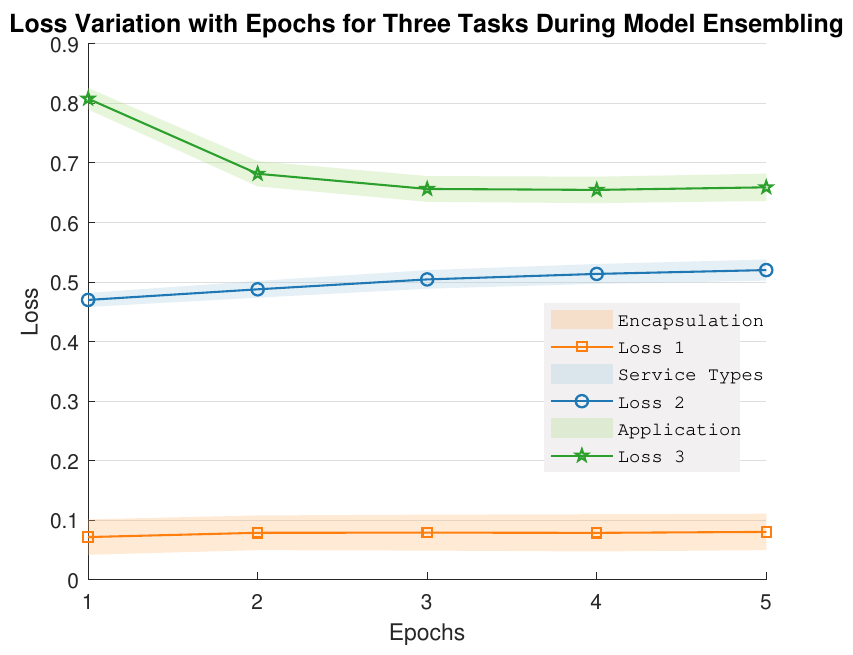} 
		\label{fig:5-1}
	\end{minipage}
	\hfill
	\begin{minipage}{0.23\textwidth} 
		\centering
		\includegraphics[width=\textwidth]{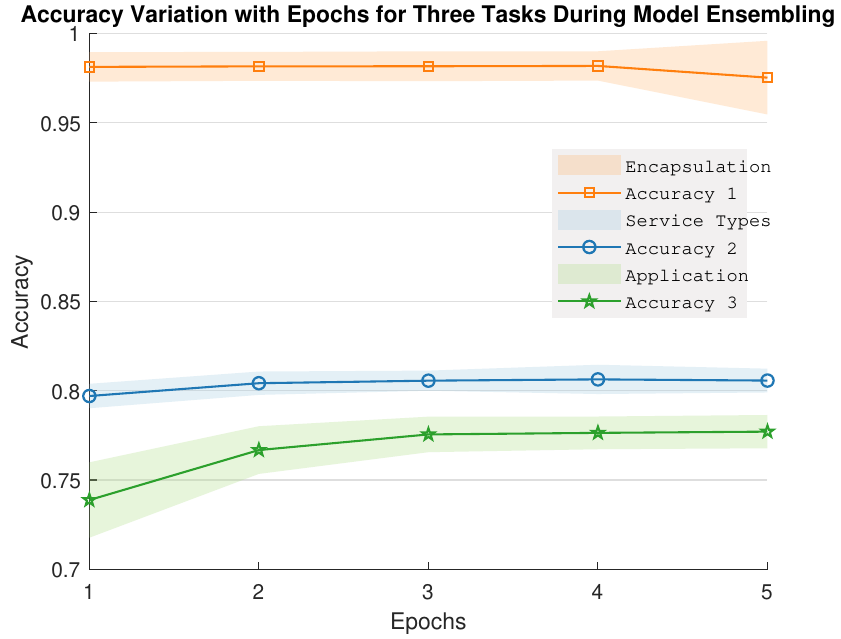} 
		\label{fig:5-2}
	\end{minipage}
	\vspace{-5mm}	
	\caption{Model extension performance in Scenarios 5.3.1}
	\label{fig:5}
	\end{figure}
	
	The accuracy and loss function values of the model across the three tasks as a function of training epochs are shown in Figure 5. The results indicate that even with a very small learning rate ($1 \times 10^{-4}$), the model can still achieve convergence within five epochs, which is quite rapid. After conducting ten repeated experiments, we found that the average accuracy of SNAKE system's models in this scenario for the three tasks reached: 98.84$\%$, 80.87$\%$, and 77.78$\%$, which are comparable to the optimal results of each independent model in Experiment 5.2, indicating that the SNAKE system has good compatibility in this scenario. In this experimental scenario, the basic settings include a batch size of 128, a tower network with only two layers and no dropout, and the parameters of the expert network are also frozen. The impact of these hyper-parameter settings on the model performance in this scenario, along with related discussions, can be found in Section 5.4.
	
	\vspace{-8pt}\subsubsection{Category Expansion Scenarios}
	In this scenario, we aim to evaluate the effect of model fusion when the number of sample categories increases in a certain classification task. In this context, we utilize the Application classification task from the ISCX-VPN-2016 and IPTAS datasets to construct a model fusion scenario with an increased number of sample categories. The specific setup of the scenario is shown in the Figure 4(b).\par
	We set the gating layer to top-k mode and redefined the newly added types. Unlike the simple default mode, configuring the gating mechanism is very important and complex in the case where multiple expert networks correspond to a single gating network. We first present the model fusion results under the top-k gating configuration, as shown in Figure 6. \par
	
		\begin{figure}[h]
		\centering
		
		\begin{minipage}{0.22\textwidth}
			\centering
			\includegraphics[width=\textwidth]{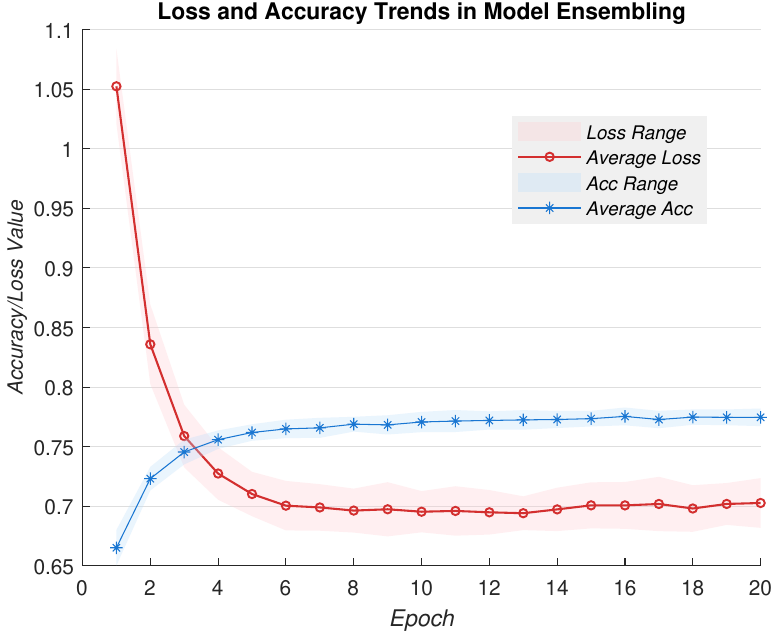} 
			\label{fig:6-1}
		\end{minipage}
		\hfill
		\begin{minipage}{0.24\textwidth}
			\centering
			\includegraphics[width=\textwidth]{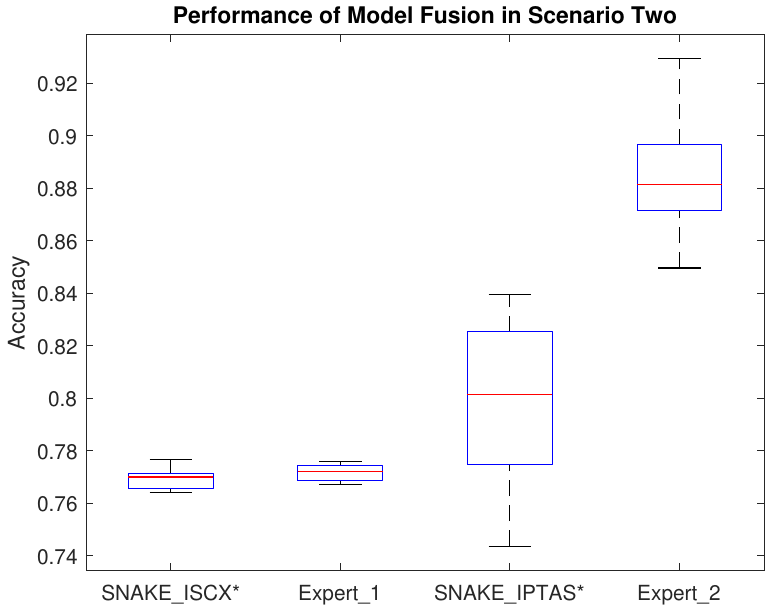} 
			\label{fig:6-2}
		\end{minipage}	
		\caption{Model extension performance in Scenarios 5.3.2}
		\label{fig:6}
		\vspace{-8pt}
	\end{figure}
	In our experiments, we found that the model fusion in this scenario is similar to incremental learning. Although the top-k gating simply averages the two outputs, which is a straightforward linear transformation, the Tower does need to relearn the relationship between features and the remapped categories. Therefore, we appropriately increased the learning rate to 0.001 for this experiment. As shown in Figure 6, it is evident that the model fusion converges in about five rounds, which is still relatively quick. We divided the final labels into two domains of the original dataset to observe their respective classification performance. Overall, the results are quite close to those of the individual models, although the classification results of IPTAS are slightly worse. However, it is possible to set hyper-parameters more reasonably to bring the accuracy closer to that of the original model; further details are discussed in Section 5.4. In addition, we also analyzed the reasons why the trainable configuration is not suitable for this case. However, to maintain the flow of the text, the details are included in \textit{Appendix Section C}.
	
	\vspace{-8pt}\subsubsection{Category Refinement Scenarios}
	
	In this scenario, after implementing the original classification task, the newly added expert model refines the classification of a specific category. We constructed this fusion scenario using the USTC-TPC2016 dataset. Here, in order to highlight the advantages of SNAKE system and the setting of trainable gates, we have established a scenario similar to incremental learning, where Expert One has acquired five categories of ten types of malicious samples, while Expert Two holds the remaining five categories. This setup closely resembles the scenario of traffic analysis in network intrusion detection, where new types of attacks are constantly emerging. The model needs to maintain its effectiveness in recognizing existing attacks while rapidly adapting to identify new malicious behavior.Due to the limited size of the dataset and the significant differences between malicious and normal behaviors, the performance of the model is quite good both before and after fusion. Here, we would like to focus primarily on the improvement in the overall recognition performance of Expert Task One after the model fusion, particularly concerning the inclusion of newly added attack samples. The experimental results are shown in Figure 7.
	
	\begin{figure}[h]
		\centering
		\includegraphics[width=0.33\textwidth]{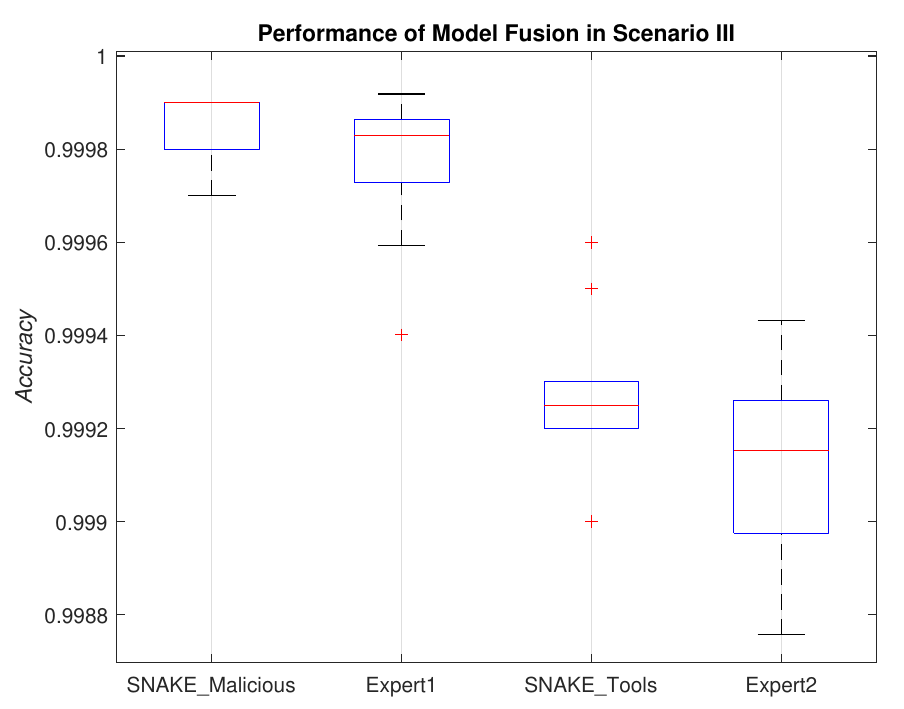} 
		\caption{Model extension performance in Scenarios 5.3.3}
		\label{fig7}
		\vspace{-8pt}
	\end{figure}
		
	The experimental results demonstrate that the SNAKE system performs the model fusion process very rapidly.We also found that this model successfully leverages the additional sample information provided by Expert Two to improve the recognition accuracy of the sample set in Task One (65$\%$ approximately), which includes new attack samples. The results clearly show the advantages of the trainable gating settings in this scenario, which resembles incremental learning and involves the refinement of categories.\par

	\begin{figure*}[h]
		\centering
		\begin{subfigure}{0.44\linewidth}
			\centering
			\includegraphics[scale=.32]{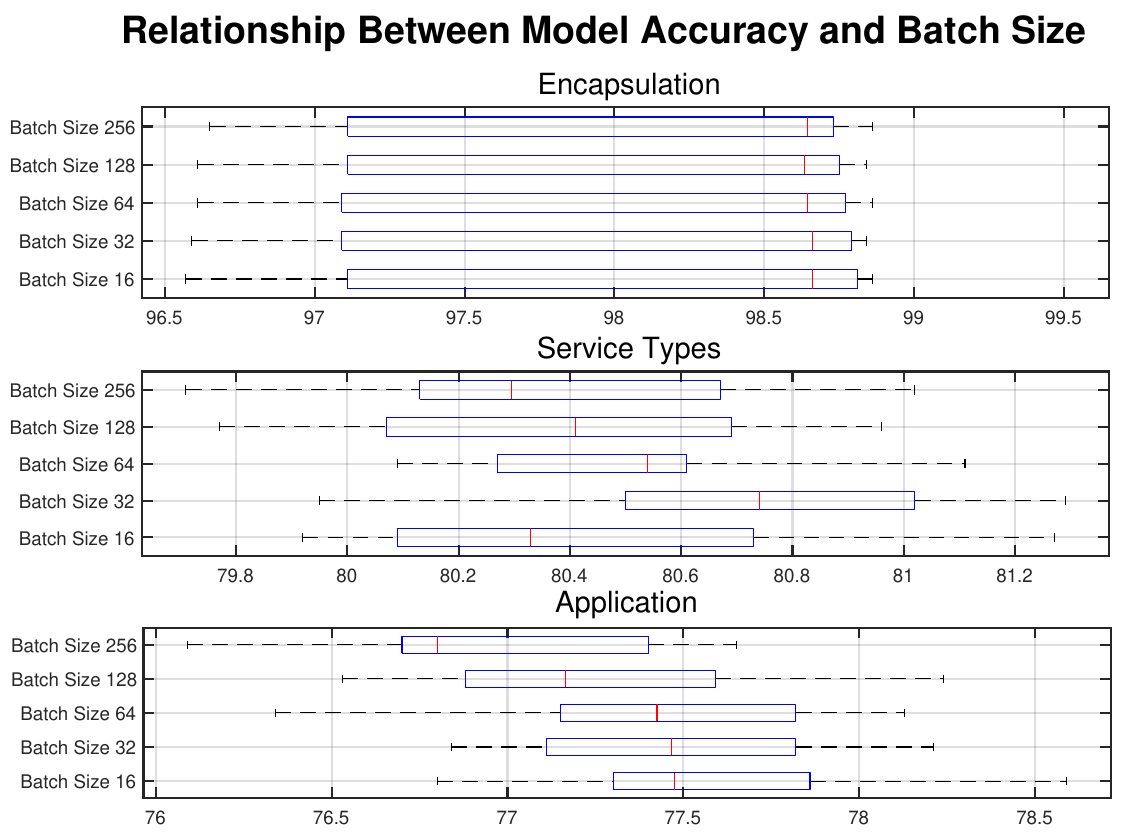}
		\end{subfigure}
		\hspace{5pt}
		\begin{subfigure}{0.44\linewidth}
			\centering
			\includegraphics[scale=.45]{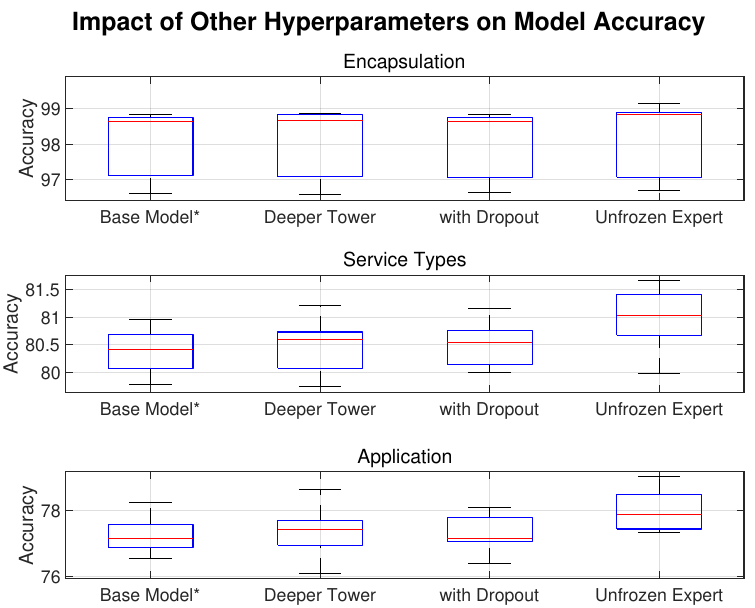}
		\end{subfigure}
		\caption{Impact of Hyper-parameter Settings on Model Performance in Task Attribute-Independent Scenarios}
		\label{figure 8}
	\end{figure*}
	
	\vspace{-8pt}\subsection{Impact of Hyper-parameter Settings on Model Performance}
	In this section of the experiment, we aimed to verify whether the settings of certain hyper-parameters in SNAKE would affect the overall performance of the model. First, we analyzed Task Attribute-Independent Scenarios and utilized three classification from the ISCX-VPN-2016 dataset to examine the impact of different batch sizes, the number of layers in the tower, and the dropout rates in the tower layers on the model's recognition performance. The corresponding trends in model accuracy are illustrated in Figure 8.\par

	From the figure, it is not difficult to see that the batch size has a relatively minor average impact on the model accuracy in this scenario. However, a smaller batch size can lead to bolder updates in the direction of optimization, potentially resulting in slightly better performance. That said, when fusing models, if there are too many label types across two or more tasks, a smaller batch size might result in some labels not being represented during a single training iteration, which could negatively affect the training process. Therefore, we believe that setting the batch size to 128 is more reasonable. In terms of other hyper-parameters, setting dropout for the Tower network results in a slight improvement in the model's average performance, although this is not very noticeable. On the other hand, increasing the depth of the Tower and not freezing the parameters of the expert models can lead to a significant enhancement in model performance. However, both of these approaches may slightly increase the computational resource requirements, so whether to implement these settings should depend on the specific needs of the user.\par
	
	For the scenario in Experiment 5.3.2, the selection of batch size is similar to the situation mentioned above, so we will not elaborate further here. In this case, the base model did not use dropout, the learning rate was set to 0.001, and the parameters of the expert networks were frozen. We conducted experiments with a smaller learning rate, deployed dropout layers in the Tower layer, and allowed the parameters of the expert networks to be updated, observing the impact of these parameter settings on the experimental accuracy. The model's classification accuracy for the IPTAS and ISCX-VPN-2016 datasets under these configurations is shown in Figure 9.\par
	
	\begin{figure}[h]
		\centering
		\includegraphics[width=0.33\textwidth]{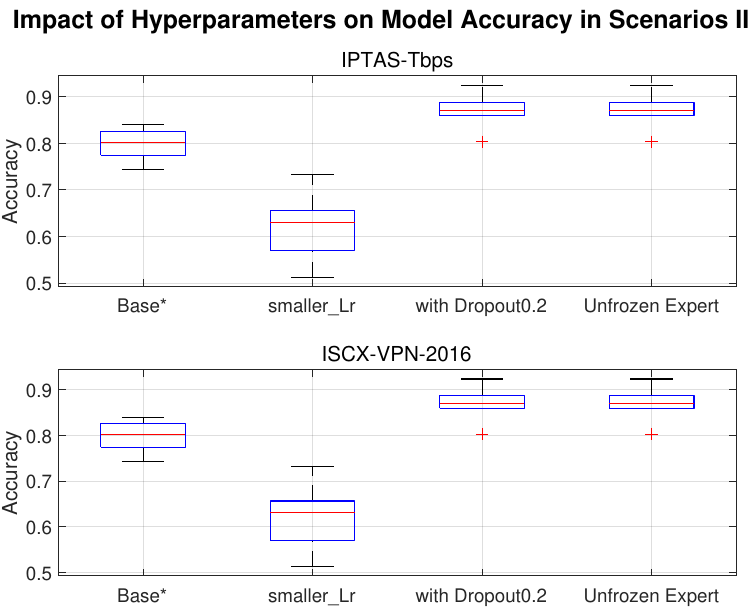} 
		\caption{Impact of Hyper-parameter Settings on Model Performance in Category Expansion Scenarios}
		\label{fig9}
		\vspace{-8pt}
	\end{figure}
	
	From the figure, we can conclude that in this scenario, model fusion indeed requires a slightly higher learning rate to quickly update the mapping relationships. Additionally, configuring dropout layers in the Tower network helps improve the model's accuracy. While allowing the expert models to be updated can enhance the performance of the model fusion, it requires more computational resources and needs to be allocated as needed. Furthermore, as we observed in Section 5.3.2, after configuring the dropout layers, the model's recognition performance has fully reached the recognition accuracy of the independent expert models. This indicates that as long as the hyper-parameters are set appropriately, the SNAKE system can effectively approximate the capabilities of individual models across each classification dimension.\par
	
	\subsection{Overall Effects of Integrating Numerous Tasks on Model Performance}
	In this final experiment, we aim to showcase the model expansion that the SNAKE system has achieved so far, as well as its recognition performance on related tasks. While there is potential to further expand upon a variety of additional models and tasks, this paper will limit its scope to the current findings. Figure 10 illustrates the detection performance on various task test sets following model fusion, in comparison to the original individual models. Due to constraints on the length of this manuscript, the training processes and detailed results of the individual models are provided in \textit{Appendix D} .\par
		\begin{figure}[h]
		\centering
		\includegraphics[width=0.33\textwidth]{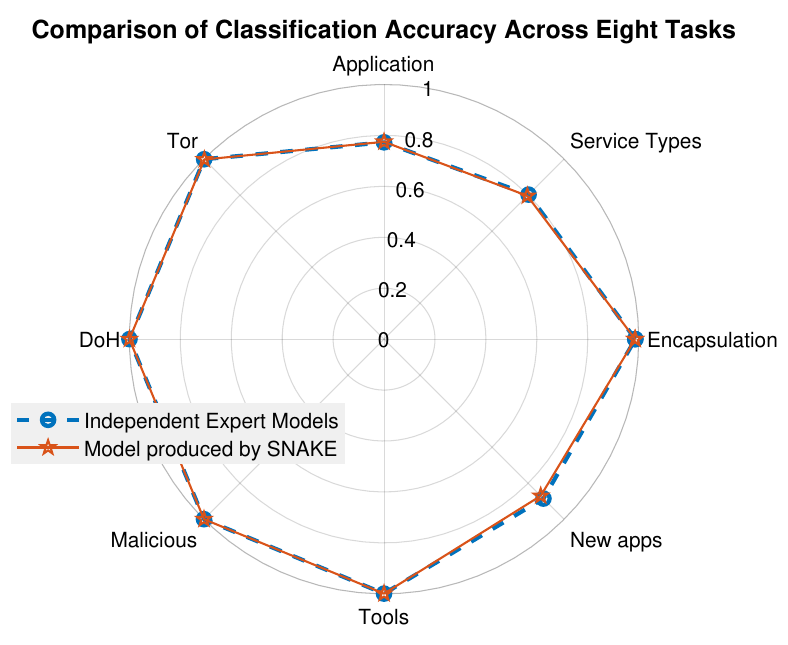} 
		\caption{Overall Effects of Integrating Numerous Tasks on Model Performance}
		\label{fig10}
		\vspace{-8pt}
	\end{figure}
	
	Based on the figure, it can be observed that the performance of the model fusion for each task closely aligns with that of the individual models, indicating that the SNAKE system has preliminarily developed a relatively universal large-scale model in the field of traffic classification.
	
	\section{Limitations and Challenges}
	After extensive experimental validation, the SNAKE system has demonstrated the ability to rapidly integrate a wide range of categories in the field of network traffic classification, exhibiting strong scalability in different scenarios and adapting well to the ever-changing network environment. However, there are still areas that can be further improved.\par
	
	First, we believe that the architecture of the SNAKE system is not yet efficient enough. If we can identify a large pre-trained model that is highly compatible with various traffic classification tasks, it could significantly enhance the SNAKE architecture, transforming it into a three-layer model. In this setup, different classification tasks could be fine-tuned using the large pre-trained model, allowing the SNAKE system to only read a small fine-tuning layer as the expert network. This would greatly reduce the complexity of the SNAKE system's operation.\par
	
	Secondly, most existing network traffic classification approaches still treat individual streams as the classification unit, which limits the information gain from applying large models. To fully leverage the benefits of high-parameter models, traffic classification should evolve towards using multiple streams over longer time windows as the recognition unit. The goal should shift from atomic classification results to summarizing network events over time periods. This technological evolution could enable the SNAKE system to transition from a sample-based mixed expert model to a token-based mixed expert model architecture, similar to contemporary mainstream large language models.\par
	These two points represent the main directions for optimizing the SNAKE system. Additionally, further exploration is needed regarding the integration of the SNAKE system with other traffic classification tasks beyond those discussed in this paper.\par
	
	\section{Conclusion}
	In this paper, we introduced the SNAKE system to address the challenges faced in traffic recognition tasks, which stem from the diverse environments in which traffic data is generated, the various input formats, classification granularity, and recognition objectives. Existing traffic recognition models often exhibit limited functionality, and the process of training numerous models and selecting them is both inefficient and difficult to coordinate.\par

	By leveraging the MMoE (Multi-gate Mixture of Experts) architecture, the SNAKE system can swiftly integrate pre-trained models from different traffic classification domains, merging them into a large-scale traffic classification model capable of performing multiple classification functions. We explored three distinct task expansion modes and examined their compatibility through three different gating layer configurations.\par

	Additionally, we tested various factors influencing the multi-recognition accuracy of the SNAKE system, including the expert model structure, different task expansion modes, hyper-parameters, and tower layer network designs. Ultimately, we conducted a comprehensive evaluation across eight traffic classification tasks using several publicly available datasets. The experimental results demonstrated that the SNAKE system achieved a commendable overall recognition accuracy, enabling efficient reasoning of multi-attribute traffic intelligence.\par

	This work not only highlights the potential of the SNAKE system in enhancing traffic classification but also sets the stage for future research aimed at further optimizing and expanding its capabilities.\par

	\newpage	
	\bibliographystyle{plain}
	\bibliography{\jobname}

\begin{thebibliography}{10}

\bibitem{2014Trends}
Jemal~H. Abawajy.
\newblock Trends in crime toolkit development.
\newblock {\em Network Security Technologies Design \& Applications}, 2014.

\bibitem{mobile}
Giuseppe Aceto, Domenico Ciuonzo, Antonio Montieri, and Antonio Pescap{\'e}.
\newblock Toward effective mobile encrypted traffic classification through deep
  learning.
\newblock {\em Neurocomputing}, 409:306--315, 2020.

\bibitem{distiller}
Giuseppe Aceto, Domenico Ciuonzo, Antonio Montieri, and Antonio Pescap{\'e}.
\newblock Distiller: Encrypted traffic classification via multimodal multitask
  deep learning.
\newblock {\em Journal of Network and Computer Applications}, 183:102985, 2021.

\bibitem{DNS}
Ahmad~Reda Alzighaibi.
\newblock Detection of doh traffic tunnels using deep learning for encrypted
  traffic classification.
\newblock {\em Computers}, 12(3):47, 2023.

\bibitem{AZAB2022}
Ahmad Azab, Mahmoud Khasawneh, Saed Alrabaee, Kim-Kwang~Raymond Choo, and Maysa
  Sarsour.
\newblock Network traffic classification: Techniques, datasets, and challenges.
\newblock {\em Digital Communications and Networks}, 2022.

\bibitem{ISP}
Ali~Javed Azhar-ud din, Ayesha Hanif, M~Awais Azam, and Tasawer Hussain.
\newblock Development of postpaid and prepaid billing system for isps.
\newblock {\em Proceedings Appeared on IOARP Digital Library}, 2016.

\bibitem{malware}
Onur Barut, Yan Luo, Peilong Li, and Tong Zhang.
\newblock R1dit: Privacy-preserving malware traffic classification with
  attention-based neural networks.
\newblock {\em IEEE Transactions on Network and Service Management},
  20(2):2071--2085, 2022.

\bibitem{2012A}
Tomasz Bujlow, Tahir Riaz, and Jens~Myrup Pedersen.
\newblock A method for classification of network traffic based on c5.0 machine
  learning algorithm.
\newblock In {\em 2012 International Conference on Computing, Networking and
  Communications (ICNC)}, 2012.

\bibitem{10.1007/978-3-662-45670-5_8}
Zigang Cao, Gang Xiong, Yong Zhao, Zhenzhen Li, and Li~Guo.
\newblock A survey on encrypted traffic classification.
\newblock In Lynn Batten, Gang Li, Wenjia Niu, and Matthew Warren, editors,
  {\em Applications and Techniques in Information Security}, pages 73--81,
  Berlin, Heidelberg, 2014. Springer Berlin Heidelberg.

\bibitem{chen2022a3c}
Zihan Chen, Guang Cheng, Ziheng Xu, Keya Xu, Yuhang Shan, and Jiakang Zhang.
\newblock A3c system: one-stop automated encrypted traffic labeled sample
  collection, construction and correlation in multi-systems.
\newblock {\em Applied Sciences}, 12(22):11731, 2022.

\bibitem{DBLP:journals/corr/abs-2204-09465}
Tianyu Cui, Gaopeng Gou, Gang Xiong, Zhen Li, Mingxin Cui, and Chang Liu.
\newblock Siamhan: Ipv6 address correlation attacks on {TLS} encrypted traffic
  via siamese heterogeneous graph attention network.
\newblock {\em CoRR}, abs/2204.09465, 2022.

\bibitem{9916060}
Jianbang Dai, Xiaolong Xu, Honghao Gao, Xinheng Wang, and Fu~Xiao.
\newblock Shape: A simultaneous header and payload encoding model for encrypted
  traffic classification.
\newblock {\em IEEE Transactions on Network and Service Management},
  20(2):1993--2012, 2023.

\bibitem{dong2024deep}
Wenqi Dong, Jing Yu, Xinjie Lin, Gaopeng Gou, and Gang Xiong.
\newblock Deep learning and pre-training technology for encrypted traffic
  classification: A comprehensive review.
\newblock {\em Neurocomputing}, page 128444, 2024.

\bibitem{DBLP:conf/uss/Dong0WLZTP0XX23}
Yutao Dong, Qing Li, Kaidong Wu, Ruoyu Li, Dan Zhao, Gareth Tyson, Junkun Peng,
  Yong Jiang, Shutao Xia, and Mingwei Xu.
\newblock Horuseye: {A} realtime iot malicious traffic detection framework
  using programmable switches.
\newblock In Joseph~A. Calandrino and Carmela Troncoso, editors, {\em 32nd
  {USENIX} Security Symposium, {USENIX} Security 2023, Anaheim, CA, USA, August
  9-11, 2023}, pages 571--588. {USENIX} Association, 2023.

\bibitem{VPN2016}
Gerard Draper-Gil, Arash~Habibi Lashkari, Mohammad Saiful~Islam Mamun, and
  Ali~A Ghorbani.
\newblock Characterization of encrypted and vpn traffic using time-related.
\newblock In {\em Proceedings of the 2nd international conference on
  information systems security and privacy (ICISSP)}, pages 407--414, 2016.

\bibitem{eigen2013learning}
David Eigen, Marc'Aurelio Ranzato, and Ilya Sutskever.
\newblock Learning factored representations in a deep mixture of experts.
\newblock {\em arXiv preprint arXiv:1312.4314}, 2013.

\bibitem{fedus2022switch}
William Fedus, Barret Zoph, and Noam Shazeer.
\newblock Switch transformers: Scaling to trillion parameter models with simple
  and efficient sparsity.
\newblock {\em Journal of Machine Learning Research}, 23(120):1--39, 2022.

\bibitem{fu2021realtime}
Chuanpu Fu, Qi~Li, Meng Shen, and Ke~Xu.
\newblock Realtime robust malicious traffic detection via frequency domain
  analysis.
\newblock In {\em Proceedings of the 2021 ACM SIGSAC Conference on Computer and
  Communications Security}, pages 3431--3446, 2021.

\bibitem{wifi}
Giuseppe Granato, Alessio Martino, Andrea Baiocchi, and Antonello Rizzi.
\newblock Graph-based multi-label classification for wifi network traffic
  analysis.
\newblock {\em Applied Sciences}, 12(21):11303, 2022.

\bibitem{vpn}
Lulu Guo, Qianqiong Wu, Shengli Liu, Ming Duan, Huijie Li, and Jianwen Sun.
\newblock Deep learning-based real-time vpn encrypted traffic identification
  methods.
\newblock {\em Journal of Real-Time Image Processing}, 17(1):103--114, 2020.

\bibitem{9343185}
Yuzong Hu, Futai Zou, Linsen Li, and Ping Yi.
\newblock Traffic classification of user behaviors in tor, i2p, zeronet,
  freenet.
\newblock In {\em 2020 IEEE 19th International Conference on Trust, Security
  and Privacy in Computing and Communications (TrustCom)}, pages 418--424,
  2020.

\bibitem{2009A}
Shijun Huang, Kai Chen, Chao Liu, Alei Liang, and Haibing Guan.
\newblock A statistical-feature-based approach to internet traffic
  classification using machine learning.
\newblock In {\em Proceedings of the International Conference on Ultra Modern
  Telecommunications, ICUMT 2009, 12-14 October 2009, St. Petersburg, Russia},
  2009.

\bibitem{jafri2024leo}
Syed~Usman Jafri, Sanjay Rao, Vishal Shrivastav, and Mohit Tawarmalani.
\newblock Leo: Online $\{$ML-based$\}$ traffic classification at
  $\{$Multi-Terabit$\}$ line rate.
\newblock In {\em 21st USENIX Symposium on Networked Systems Design and
  Implementation (NSDI 24)}, pages 1573--1591, 2024.

\bibitem{ddos}
Danial Javaheri, Saeid Gorgin, Jeong-A Lee, and Mohammad Masdari.
\newblock Fuzzy logic-based ddos attacks and network traffic anomaly detection
  methods: Classification, overview, and future perspectives.
\newblock {\em Information Sciences}, 626:315--338, 2023.

\bibitem{2004Transport}
Thomas Karagiannis, Andre Broido, Michalis Faloutsos, and Kc~Claffy.
\newblock Transport layer identification of p2p traffic.
\newblock In {\em ACM SIGCOMM conference on Internet measurement}, 2004.

\bibitem{khanna2006application}
Gunjan Khanna, Kirk Beaty, Gautam Kar, and Andrzej Kochut.
\newblock Application performance management in virtualized server
  environments.
\newblock In {\em 2006 IEEE/IFIP Network Operations and Management Symposium
  NOMS 2006}, pages 373--381. IEEE, 2006.

\bibitem{iot}
Rakesh Kumar, Mayank Swarnkar, Gaurav Singal, and Neeraj Kumar.
\newblock Iot network traffic classification using machine learning algorithms:
  An experimental analysis.
\newblock {\em IEEE Internet of Things Journal}, 9(2):989--1008, 2021.

\bibitem{Tor2016}
Arash~Habibi Lashkari, Gerard~Draper Gil, Mohammad Saiful~Islam Mamun, and
  Ali~A Ghorbani.
\newblock Characterization of tor traffic using time based features.
\newblock In {\em International Conference on Information Systems Security and
  Privacy}, volume~2, pages 253--262. SciTePress, 2017.

\bibitem{li2024hierarchical}
Weikai Li, Ding Wang, Zijian Ding, Atefeh Sohrabizadeh, Zongyue Qin, Jason
  Cong, and Yizhou Sun.
\newblock Hierarchical mixture of experts: Generalizable learning for
  high-level synthesis.
\newblock {\em arXiv preprint arXiv:2410.19225}, 2024.

\bibitem{DBLP:conf/sp/LiWDL24}
Xiang Li, Dashuai Wu, Haixin Duan, and Qi~Li.
\newblock Dnsbomb: {A} new practical-and-powerful pulsing dos attack exploiting
  {DNS} queries-and-responses.
\newblock In {\em {IEEE} Symposium on Security and Privacy, {SP} 2024, San
  Francisco, CA, USA, May 19-23, 2024}, pages 4478--4496. {IEEE}, 2024.

\bibitem{lin2022bert}
Xinjie Lin, Gang Xiong, Gaopeng Gou, Zhen Li, Junzheng Shi, and Jing Yu.
\newblock Et-bert: A contextualized datagram representation with pre-training
  transformers for encrypted traffic classification.
\newblock In {\em Proceedings of the ACM Web Conference 2022}, pages 633--642,
  2022.

\bibitem{FS-net}
Chang Liu, Longtao He, Gang Xiong, Zigang Cao, and Zhen Li.
\newblock Fs-net: A flow sequence network for encrypted traffic classification.
\newblock In {\em IEEE INFOCOM 2019 - IEEE Conference on Computer
  Communications}, pages 1171--1179, 2019.

\bibitem{20243917120586}
Tiantian Liu, Feng Lin, Zhongjie Ba, Li~Lu, Zhan Qin, and Kui Ren.
\newblock Micguard: A comprehensive detection system against out-of-band
  injection attacks for different level microphone-based devices.
\newblock pages 3963 -- 3978, Philadelphia, PA, United states, 2024.
\newblock Detection models;Detection system;Electromagnetic attack;Injected
  signal;Input interface;Intelligent applications;Out of band;Passive detection
  systems;Prior information;Public concern;.

\bibitem{lopez2017network}
Manuel Lopez-Martin, Belen Carro, Antonio Sanchez-Esguevillas, and Jaime
  Lloret.
\newblock Network traffic classifier with convolutional and recurrent neural
  networks for internet of things.
\newblock {\em IEEE access}, 5:18042--18050, 2017.

\bibitem{algorithm1}
Manuel Lopez-Martin, Belen Carro, Antonio Sanchez-Esguevillas, and Jaime
  Lloret.
\newblock Network traffic classifier with convolutional and recurrent neural
  networks for internet of things.
\newblock {\em IEEE Access}, 5:18042--18050, 2017.

\bibitem{ma2018modeling}
Jiaqi Ma, Zhe Zhao, Xinyang Yi, Jilin Chen, Lichan Hong, and Ed~H Chi.
\newblock Modeling task relationships in multi-task learning with multi-gate
  mixture-of-experts.
\newblock In {\em Proceedings of the 24th ACM SIGKDD international conference
  on knowledge discovery \& data mining}, pages 1930--1939, 2018.

\bibitem{DoH}
Mohammadreza MontazeriShatoori, Logan Davidson, Gurdip Kaur, and Arash~Habibi
  Lashkari.
\newblock Detection of doh tunnels using time-series classification of
  encrypted traffic.
\newblock {\em 2020 IEEE Intl Conf on Dependable, Autonomic and Secure
  Computing, Intl Conf on Pervasive Intelligence and Computing, Intl Conf on
  Cloud and Big Data Computing, Intl Conf on Cyber Science and Technology
  Congress (DASC/PiCom/CBDCom/CyberSciTech)}, pages 63--70, 2020.

\bibitem{parisi2019continual}
German~I Parisi, Ronald Kemker, Jose~L Part, Christopher Kanan, and Stefan
  Wermter.
\newblock Continual lifelong learning with neural networks: A review.
\newblock {\em Neural networks}, 113:54--71, 2019.

\bibitem{DBLP:conf/ndss/QingYDCL000024}
Yuqi Qing, Qilei Yin, Xinhao Deng, Yihao Chen, Zhuotao Liu, Kun Sun, Ke~Xu, Jia
  Zhang, and Qi~Li.
\newblock Low-quality training data only? {A} robust framework for detecting
  encrypted malicious network traffic.
\newblock In {\em 31st Annual Network and Distributed System Security
  Symposium, {NDSS} 2024, San Diego, California, USA, February 26 - March 1,
  2024}. The Internet Society, 2024.

\bibitem{algorithm3}
Shahbaz Rezaei and Xin Liu.
\newblock Multitask learning for network traffic classification.
\newblock In {\em 2020 29th International Conference on Computer Communications
  and Networks (ICCCN)}, pages 1--9, 2020.

\bibitem{tor}
Debmalya Sarkar, P~Vinod, and Suleiman~Y Yerima.
\newblock Detection of tor traffic using deep learning.
\newblock In {\em 2020 IEEE/ACS 17th International Conference on Computer
  Systems and Applications (AICCSA)}, pages 1--8. IEEE, 2020.

\bibitem{NPB}
GVK Sasirekha, GH~Annapoorna, Madhav Rao, Jyotsna Bapat, and Debabrata Das.
\newblock Ml-augmented network packet broker based anomaly detection at
  iiot-edge egress port.
\newblock In {\em 2023 IEEE International Conference on Advanced Networks and
  Telecommunications Systems (ANTS)}, pages 1--6. IEEE, 2023.

\bibitem{algorithm5}
Haifeng Sun, Yunming Xiao, Jing Wang, Jingyu Wang, Qi~Qi, Jianxin Liao, and
  Xiulei Liu.
\newblock Common knowledge based and one-shot learning enabled multi-task
  traffic classification.
\newblock {\em IEEE Access}, 7:39485--39495, 2019.

\bibitem{report2023}
International~Telecommunication Union.
\newblock Measuring digital development: Facts and figures 2023, 2023.

\bibitem{van2022three}
Gido~M Van~de Ven, Tinne Tuytelaars, and Andreas~S Tolias.
\newblock Three types of incremental learning.
\newblock {\em Nature Machine Intelligence}, 4(12):1185--1197, 2022.

\bibitem{7899588}
Wei Wang, Ming Zhu, Xuewen Zeng, Xiaozhou Ye, and Yiqiang Sheng.
\newblock Malware traffic classification using convolutional neural network for
  representation learning.
\newblock In {\em 2017 International Conference on Information Networking
  (ICOIN)}, pages 712--717, 2017.

\bibitem{wang2017malware}
Wei Wang, Ming Zhu, Xuewen Zeng, Xiaozhou Ye, and Yiqiang Sheng.
\newblock Malware traffic classification using convolutional neural network for
  representation learning.
\newblock In {\em 2017 International conference on information networking
  (ICOIN)}, pages 712--717. IEEE, 2017.

\bibitem{wang2020id}
Xiaoliang Wang, Ke~Xu, Wenlong Chen, Qi~Li, Meng Shen, and Bo~Wu.
\newblock Id-based sdn for the internet of things.
\newblock {\em IEEE Network}, 34(4):76--83, 2020.

\bibitem{2010Automatic}
Yu~Wang, Yang Xiang, and Shun~Zheng Yu.
\newblock Automatic application signature construction from unknown traffic.
\newblock In {\em IEEE International Conference on Advanced Information
  Networking \& Applications}, 2010.

\bibitem{10.1145/3405672.3409492}
Lu~Xu, Daihui Dou, and H.~Jonathan Chao.
\newblock Etcnet: Encrypted traffic classification using siamese convolutional
  networks.
\newblock In {\em Proceedings of the Workshop on Network Application
  Integration/CoDesign}, NAI '20, page 51–53, New York, NY, USA, 2020.
  Association for Computing Machinery.

\bibitem{zhang2024trafficgpt}
Siyao Zhang, Daocheng Fu, Wenzhe Liang, Zhao Zhang, Bin Yu, Pinlong Cai, and
  Baozhen Yao.
\newblock Trafficgpt: Viewing, processing and interacting with traffic
  foundation models.
\newblock {\em Transport Policy}, 150:95--105, 2024.

\bibitem{zhang2021survey}
Yu~Zhang and Qiang Yang.
\newblock A survey on multi-task learning.
\newblock {\em IEEE transactions on knowledge and data engineering},
  34(12):5586--5609, 2021.

\bibitem{zhao2023survey}
Wayne~Xin Zhao, Kun Zhou, Junyi Li, Tianyi Tang, Xiaolei Wang, Yupeng Hou,
  Yingqian Min, Beichen Zhang, Junjie Zhang, Zican Dong, et~al.
\newblock A survey of large language models.
\newblock {\em arXiv preprint arXiv:2303.18223}, 2023.

\bibitem{DBLP:conf/ccs/ZhaoD0LL0024}
Xiyuan Zhao, Xinhao Deng, Qi~Li, Yunpeng Liu, Zhuotao Liu, Kun Sun, and Ke~Xu.
\newblock Towards fine-grained webpage fingerprinting at scale.
\newblock In Bo~Luo, Xiaojing Liao, Jun Xu, Engin Kirda, and David Lie,
  editors, {\em Proceedings of the 2024 on {ACM} {SIGSAC} Conference on
  Computer and Communications Security, {CCS} 2024, Salt Lake City, UT, USA,
  October 14-18, 2024}, pages 423--436. {ACM}, 2024.

\bibitem{algorithm4}
Ying Zhao, Junjun Chen, Di~Wu, Jian Teng, and Shui Yu.
\newblock Multi-task network anomaly detection using federated learning.
\newblock In {\em the Tenth International Symposium}, 2019.

\end{thebibliography}
	
	\clearpage 
	\onecolumn 
	\section*{Ethics Considerations}
	In our network traffic analysis experiments, we are committed to upholding the highest ethical standards. All datasets utilized in our research are open-source and publicly available, sourced from reputable repositories to ensure transparency and accessibility. We explicitly affirm that our study does not involve any analysis of personal or private data, thereby safeguarding individual privacy rights. Furthermore, all malicious traffic samples employed in our experiments are strictly confined to a controlled local environment managed by the researchers, ensuring that no disruptions or harm can occur to external networks or individuals.\par
	Appropriate measures were implemented to safeguard the privacy of personal information. No personal data that could potentially identify participants will be disclosed or made public. In the research reports, all data have been anonymized and securely stored on password-protected servers. Access to the raw data is restricted to members of the research team, and any identifying information was removed prior to analysis. During the project design phase, we conducted a thorough assessment of potential risks associated with the research and implemented suitable measures to mitigate or manage these risks. We ensure that participants will not experience any physical or psychological harm as a result of their participation in this study. We are committed to maintaining the legality and transparency of data usage, ensuring the accurate application and interpretation of research data. Efforts have been made to minimize the risk of data misinterpretation and misuse, and data will be utilized solely for research purposes.
	
	\section*{Compliance with Open Science Policy}
	In alignment with the principles of open science, we are committed to promoting transparency and reproducibility in our research.\par
	The data that support the findings of this study were derived in the public domain: \\
	The ISCXVPN2016 dataset is available at https://doi.org/10.5220/0005740704070414. The ISCXTor2016 dataset is available at https://doi.org/10.1109/NOMS.2006.1687567. The CIC-DoHBrw-2020 dataset is available at https://doi.org/10.1109/DASC-PICom-CBDCom-CyberSciTech49142.2020.00026. The USTC-TFC2016 dataset is available at https://doi.org/10.1109/ICOIN.2017.7899588.\par
	
	In this study, we also commit to adhering to open science principles to ensure the transparency and reproducibility of our research findings. Below is a comprehensive list of all artifacts related to this paper, along with their availability status:

	\begin{enumerate}
	\item \textbf{Source Code}:
	\begin{itemize}
		\item \textbf{Artifact Name}: SNAKE System
		\item \textbf{Description}: Source code implementing the network traffic analysis system we proposed.
		\item \textbf{Availability}: Will be made publicly available upon acceptance of the paper, hosted on GitHub.
	\end{itemize}
	
	\item \textbf{Datasets}:
	\begin{itemize}
		\item \textbf{Note}: We use only publicly available datasets, and we are happy to provide the complete process of how we handled this data, as well as our final segmentation method.
	\end{itemize}
	\item \textbf{Analysis Scripts}:
	\begin{itemize}
		\item \textbf{Artifact Name}: Analyze Script
		\item \textbf{Description}: Scripts used for processing and analyzing the experimental results.
		\item \textbf{Availability}: Will be made publicly available upon acceptance of the paper.
	\end{itemize}
		\item \textbf{Models}:
	\begin{itemize}
		\item \textbf{Artifact Name}: Models for Network Traffic Analysis
		\item \textbf{Description}: The final fused model and the individual classification task models produced in this article can provide both structure and parameters.
		\item \textbf{Availability}: Will be made publicly available upon acceptance of the paper.
	\end{itemize}
	\item \textbf{Non-Public Artifacts}:
	\begin{itemize}
		\item \textbf{The code, algorithms, models, and data in this article are all available for open-source access.}
	\end{itemize}
	\end{enumerate}
	We have made every effort to identify and list all necessary artifacts to ensure the reproducibility of our research findings. Should the availability of any artifacts change, we will update this statement accordingly.

	\twocolumn 
	\clearpage 

	\newpage 
	\appendix
	\begin{center}
		\LARGE\textbf{Appendix}
	\end{center}
	
	\section{Proof of Model Convergence}	
	This section provides the detailed derivation and proof of model convergence that were not presented in Section 4.\par
	For observing whether the model training gradually approaches the global optimal solution, it is assumed that after t rounds of supervised training following $\omega^{t+1} = \omega^t - \alpha \nabla L_o(\omega^t)$, the gap between $L_o(\omega^t)$ and the optimal value $L_o(\omega^*)$ is $\delta^t$. The following corollary can be derived from Lemma 2 in section 4:
	\begin{flalign*}
		\delta^t &= L_o(\omega^t)- L_o(\omega^*) \\
		&\leq \nabla L_o(\omega^*)^\top (\omega^t-\omega^*) \\
		&\leq \parallel \nabla L_o(\omega^*) \parallel \parallel (\omega^t-\omega^*)\parallel
	\end{flalign*}
	Therefore, we can derive the following inequality:
	\begin{equation*}
		\resizebox{0.55\hsize}{!}{$
			\frac{\delta^t}{\parallel (\omega^t-\omega^*)\parallel} \leq \parallel \nabla L_o(\omega^*) \parallel \leq \parallel \nabla L_o(\omega^t) \parallel$}
	\end{equation*} 
	Besides, the following formula can be derived based on Lemma 1:
	\begin{flalign*}
		\delta^{t+1}-\delta^t &= L_o(\omega^{t+1}-\omega^t)\\
		&\leq \nabla L_o(\omega^t)^\top (\omega^{t+1}-\omega^t) + \frac{c}{2} \parallel \omega^{t+1}-\omega^t\parallel^2_2 \\
		&= -\alpha \nabla L_o(\omega^t)^ \top \nabla L_o(\omega^t) + \frac{c\alpha^2}{2} \parallel \nabla L_o(\omega^t) \parallel^2_2\\
		&=-\alpha(1-\frac{c\alpha}{2}) \parallel \nabla L_o(\omega^t) \parallel^2_2\\
		&\leq -\alpha(1-\frac{c\alpha}{2})(\frac{\delta^t}{\parallel(\omega^t-\omega^*)\parallel})^2\\
		&\leq -z\alpha(1-\frac{c\alpha}{2}){\delta^t}^2, z = \underset{t \in [0,T]}{min} \frac{1}{\parallel(\omega^t-\omega^*)\parallel^2}
	\end{flalign*}
	It is beneficial to set the learning rate $\alpha$ here to a value below $\frac{1}{c}$, the following two conditions can be obtained from this: $ L_o(\omega^{t+1})-L_o(\omega^t) \leq 0 $ and $\frac{\delta^t}{\delta^{t+1}} \geq 1 $. Thus, we can further rewrite the corollary of Lemma 1 as the following inequality:
	\begin{equation*}
		\resizebox{0.55\hsize}{!}{$
			\frac{1}{\delta^{t+1}}-\frac{1}{\delta^{t}} \geq \frac{z\alpha (1-\frac{c\alpha}{2})\delta^t}{\delta^{t+1}}\geq z\alpha (1-\frac{c\alpha}{2})$}
	\end{equation*} 
	By summing all the inequalities over training period $t \in [0, T]$, we can obtain:
	\begin{equation*}
		\resizebox{0.45\hsize}{!}{$
			\sum_{ti=0}^T(\frac{1}{\delta^t}-\frac{1}{\delta^{t-1}}) \geq Tz\alpha(1-\frac{c\alpha}{2})$}
	\end{equation*} 
	Finally, the following formula can be derived:
	\begin{equation*}
		\resizebox{0.9\hsize}{!}{$
			\frac{1}{L_o(\omega^T)-L_o(\omega^*)} =\frac{1}{\delta^T} \geq \frac{1}{\delta^T}-\frac{1}{\delta^0}
			= \sum_{ti=0}^T(\frac{1}{\delta^t}-\frac{1}{\delta^{t-1}})
			\geq Tz\alpha(1-\frac{c\alpha}{2})$}
	\end{equation*} 
	The above derivation proves that we can reasonably set the learning rate for the overall loss function to approximate its minimum value. After $T$ rounds of training, setting learning rate $\alpha \leq \frac{1}{c}$,  the convergence upper bound of $L_o(\omega)$ can be formulated as follows:
	\begin{equation*}
		\resizebox{0.45\hsize}{!}{$ L_o(\omega^T)-L_o(\omega^*) \leq \frac{1}{Tz\alpha(1-\frac{c\alpha}{2})}
			$}
	\end{equation*} 
	This indicates that as the number of training iterations increases, $L_o(\omega^T)$ will gradually converge to the global minimum $L_o(\omega^*)$. \par
	
	\section{Detailed Description of the Dataset}
	In this section, we present detailed information about the dataset used in this paper:\par
	\textbf{ISCXVPN2016~\cite{VPN2016}:} The dataset aims to generate a representative sample of real-world traffic based on ISCX by defining a set of tasks that ensure a rich diversity and quantity. User accounts for Alice and Bob were created to utilize services such as Skype and Facebook. The dataset encompasses various types of traffic and applications, resulting in a total of 14 traffic categories, including VOIP, VPN-VOIP, P2P, and VPN-P2P, capturing both regular sessions and sessions over VPN. This diversity allows for a comprehensive reflection of actual application scenarios within the network environment. In this study, three labeling schemes were employed using the dataset: 1) VPN-nonVPN; 2) Service type, which includes Chat, Email, VoIP, File Transfer, Streaming, and P2P; and 3) 15 specific applications, including AIM, Email, Facebook, FTPS, Hangout, ICQ, Netflix, SCP, SFTP, Skype, Spotify, Vimeo, VoIPbuster, YouTube, and BitTorrent.\par
	
	\textbf{ISCXTor2016~\cite{Tor2016}:} To ensure the quantity and diversity of the dataset in the CIC, a set of tasks was defined to generate a representative sample of real-world traffic. Three user accounts were created for the collection of browser traffic, while two additional user accounts were established for communication activities, including chat, email, FTP, and P2P. For non-Tor traffic, previously collected benign traffic from a VPN project was utilized, and the Tor traffic was categorized into seven distinct traffic types. This structured approach enables the dataset to accurately reflect real-world network behavior with comprehensive and varied data.\par
	
	\textbf{CIRA-CIC-DoHBrw-2020~\cite{DoH}:} The CIRA-CIC-DoHBrw-2020 dataset employs a two-layered approach to capture benign and malicious DNS over HTTPS (DoH) traffic, as well as non-DoH traffic. To create a representative dataset, HTTPS traffic (including benign DoH and non-DoH) and DoH traffic were generated by accessing the top 10,000 Alexa websites, utilizing browsers and DNS tunneling tools that support the DoH protocol. In the first layer, the captured traffic is classified as DoH and non-DoH using a statistical features classifier. In the second layer, the DoH traffic is further characterized as benign DoH or malicious DoH through the application of a time-series classifier.\par
	
	\textbf{USTC-TFC2016~\cite{7899588}:} This dataset is divided into two parts, as detailed in Tables I and II. Part I includes ten types of malware traffic collected from public websites by CTU researchers in real network environments between 2011 and 2015. For larger traffic samples, only a subset was utilized, while smaller traffic samples were merged if they originated from the same application. Part II consists of ten types of normal traffic gathered using IXIA BPS, a professional network traffic simulation tool. Further details regarding the simulation methods can be found on the product's website. To encompass a broader range of traffic types, the ten traffic types include eight classes of common applications. The USTC-TFC2016 dataset has a total size of 3.71GB and is formatted in pcap. \par
	
	\textbf{IPTAS-Tbps~\cite{chen2022a3c}:} IPTAS-Tbps consists of traffic collected during normal webbrowsing under CERNET (China Education and Research Network). There are seven mainstream apps-running on user terminal such as JD, Sohu, and NetEase. The various transmission protocols and application versions in this data set are more up-to-date and better suited to the current network environment. \par
	
	\section {Case Analysis of Trainable Gate Configuration in Experiments 5.3.2}
	Initially, we also used trainable gated networks for Category Expansion Scenarios, but we found that there was a certain probability of encountering training anomalies. Here, we conduct a case analysis using the same data scenario as in Section 5.3.2, focusing on the results of ten experiments under trainable configurations with a learning rate of $1 \times 10^{-4}$.The conditions of these experiments are shown in Figure 10.\par
	
	\begin{figure}[h]
		\centering	
		\begin{minipage}{0.33\textwidth}
			\centering
			\includegraphics[width=\textwidth]{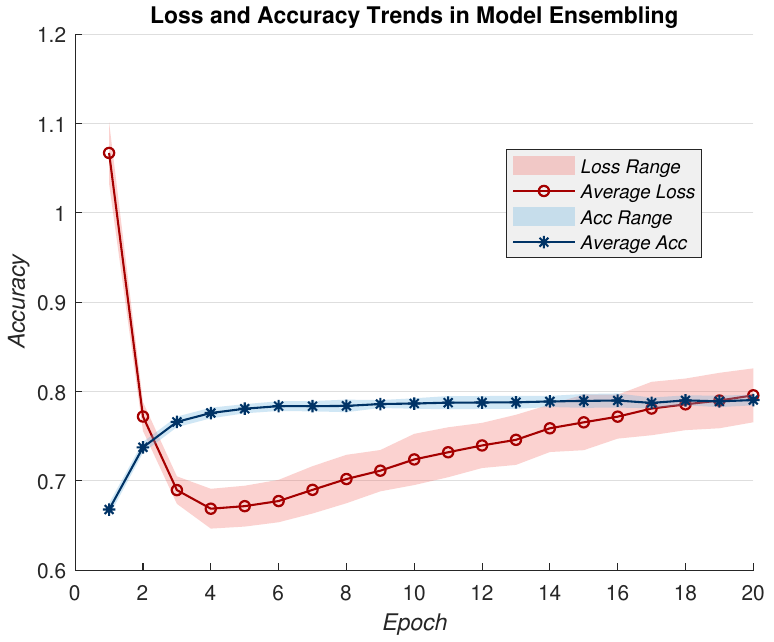} 
			\label{fig:10-1}
		\end{minipage}
		\hfill
		\begin{minipage}{0.33\textwidth}
			\centering
			\includegraphics[width=\textwidth]{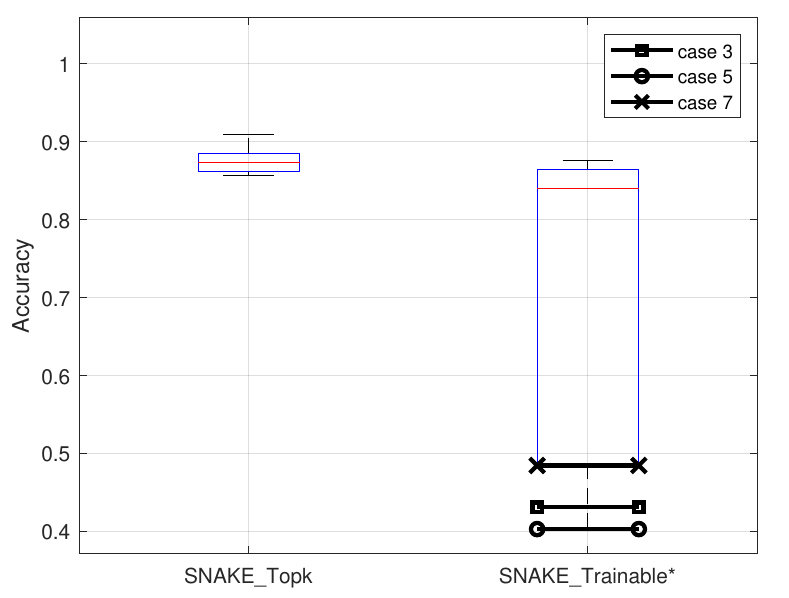} 
			\label{fig:10-2}
		\end{minipage}	
		\caption{Case Analysis of Trainable Gate Configuration}
		\label{fig:10}
	\end{figure}
	
	The upper half of the sub-figure in Figure 10 illustrates an abnormal situation with the trainable gating settings during model training. After four iterations, the model's loss function value increased instead of decreasing. Although the overall accuracy of the model did not decline correspondingly, there were actually some relatively hidden anomalies present. We subsequently grouped the sources of the identified targets and found a significant difference in recognition accuracy for the IPTAS dataset between the trainable settings and the top-k settings. Upon closer examination, we observed that in three experimental cases (case 3, 5, and 7) with trainable settings, the corresponding recognition accuracy was abnormally low. Considering the trend of the loss function, we can also conclude that this low accuracy is irretrievable.\par
	
	This situation occasionally occurs under different hyper-parameter settings, affecting the overall performance of the model ensemble. We analyze that this is due to multiple expert models sharing the Tower network base in Category Expansion Scenarios, making the model training susceptible to the influence of data distribution. These anomalies arise because the trainable gating assigns excessively high weights to the tasks of certain experts, while the relatively small data volume from other experts makes it difficult to recover. Our preliminary experiments suggest that increasing the learning rate, setting a smaller batch size, or increasing the momentum of the optimizer may alleviate this phenomenon. However, we believe that in such cases, it is safer to choose the top-k setting, even though the trainable setting may occasionally yield better results. This is because the integration of such tasks involves many unstable factors, such as sample imbalance and uneven label distribution, which can trigger these anomalies. Compared to general neural network layers, the gating layers are relatively small and fragile. Therefore, we recommend enabling the trainable gating mode only when it is very clear that the patterns of the two tasks correspond to Category Expansion Scenarios.
	\begin{figure*}[h]
		\centering
		\begin{minipage}{0.3\linewidth}
			\centering
			\includegraphics[width=\linewidth]{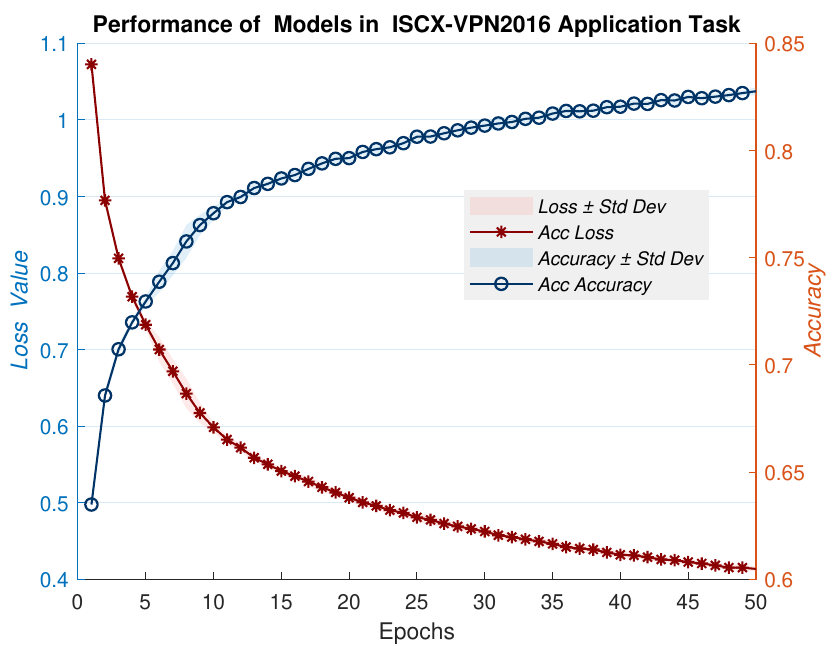}
			\label{fig:sub1}
		\end{minipage}%
		\begin{minipage}{0.3\linewidth}
			\centering
			\includegraphics[width=\linewidth]{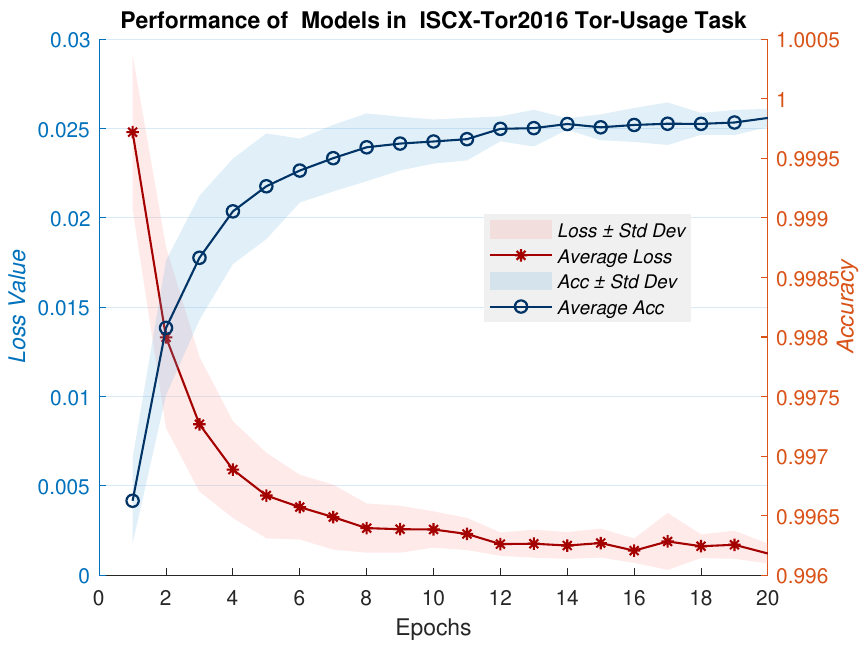}
			\label{fig:sub2}
		\end{minipage}%
		\begin{minipage}{0.3\linewidth}
			\centering
			\includegraphics[width=\linewidth]{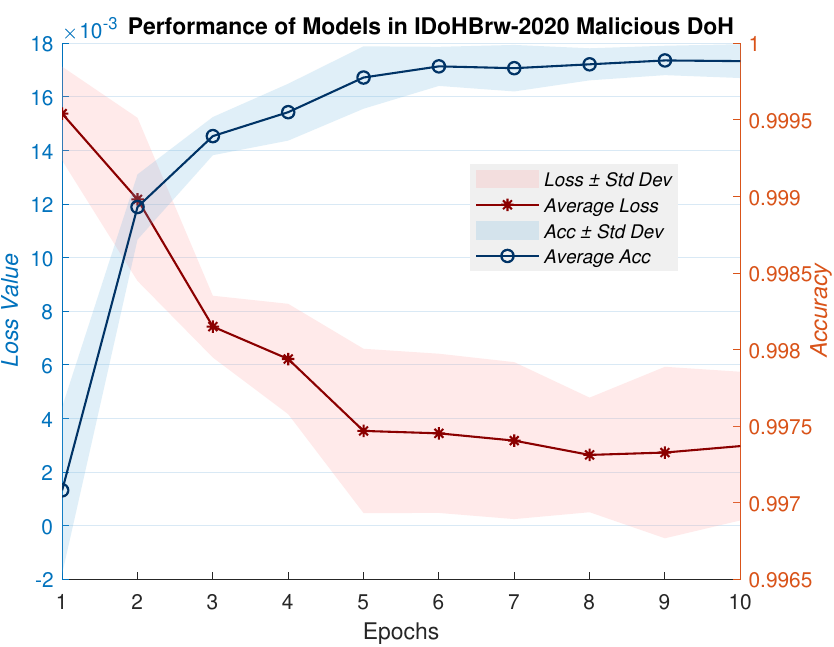}
			\label{fig:sub3}
		\end{minipage}
		
		\vspace{0.3cm} 
		
		\begin{minipage}{0.3\linewidth}
			\centering
			\includegraphics[width=\linewidth]{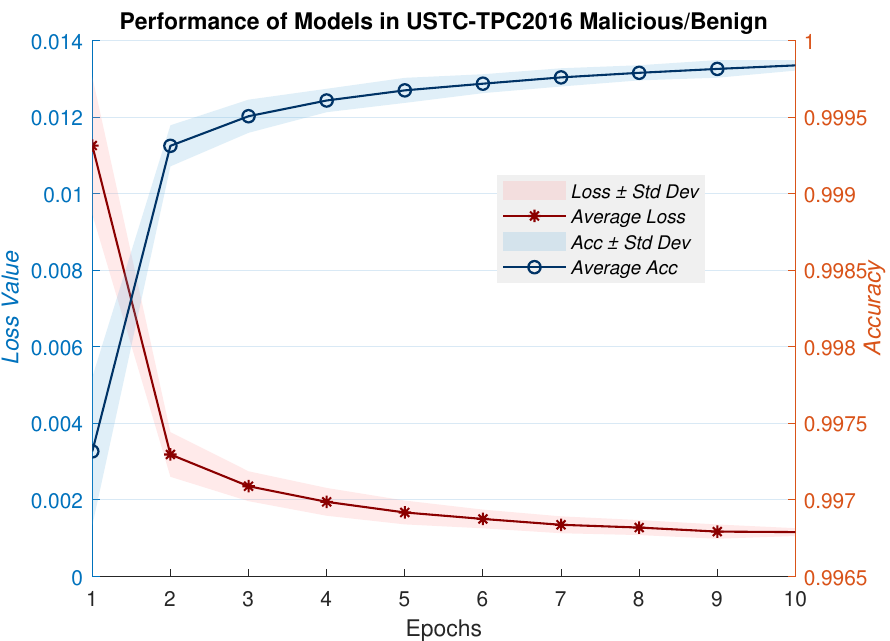}
			\label{fig:sub4}
		\end{minipage}%
		\begin{minipage}{0.3\linewidth}
			\centering
			\includegraphics[width=\linewidth]{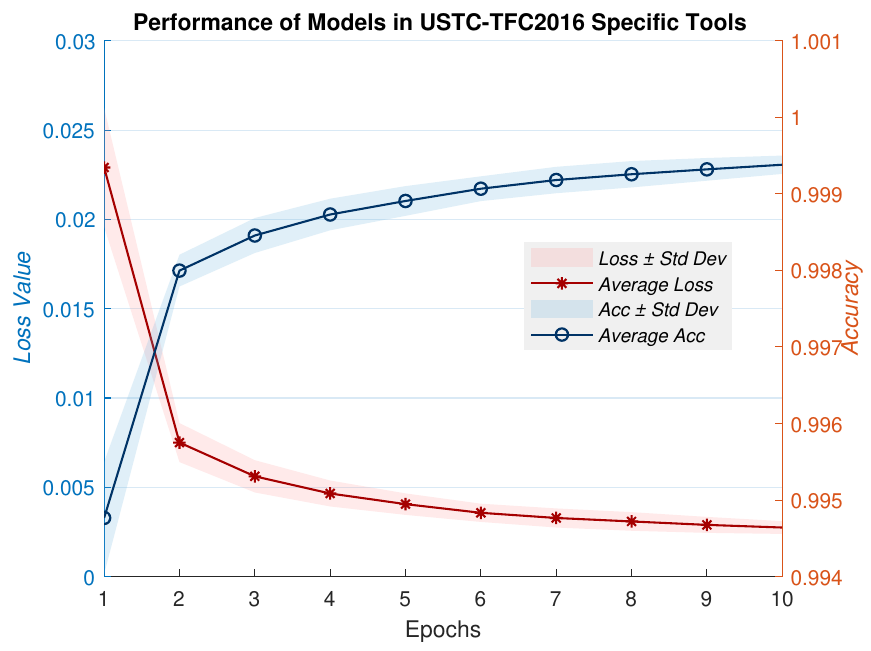}
			\label{fig:sub5}
		\end{minipage}%
		\begin{minipage}{0.3\linewidth}
			\centering
			\includegraphics[width=0.99\linewidth]{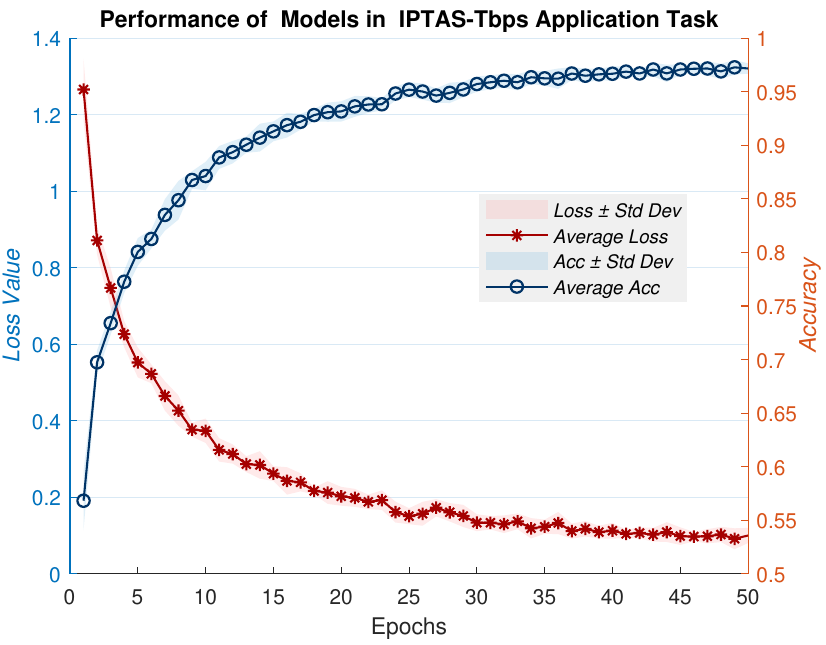}
			\label{fig:sub6}
		\end{minipage}
		
		\caption{Training Details of Each Independent Model}
		\label{fig:overall}
	\end{figure*}	
	\section{Training Processes and Detailed Results of Individual Models}
	
	This section presents the training processes and final results of various expert models read and integrated by the SNAKE system, each applied to their respective independent tasks. With the exception of the CIRA-CIC-DoHBrw-2020 dataset, which was limited to a random sampling of 7.92GB of data for training and testing due to equipment constraints, all other datasets were fully utilized and split into training, validation, and testing sets in a ratio of 75$\%$ 10$\%$, and 15$\%$. The data preprocessing and model architectures were consistent across all models; further details can be found in sections 3.4 and 3.5. For the USTC-TPC2016, CIRA-CIC-DoHBrw-2020 and ISCX-Tor datasets, we applied early stopping, training only for 20 and 10 epochs respectively, while all other models were trained for a consistent 50 epochs. The details of the training loss functions and accuracy trends for each model across epochs are illustrated in Figure 12.\par

	\begin{table}[h]
	\centering
	\caption{Classification Performance of Independent Models for Each Task from Different Datasets}
	\scalebox{0.62}{ 
		\begin{tabular}{@{}llccc@{}}
		\toprule
		\textbf{Dataset Name} & \textbf{Task Objective} & \textbf{Accuracy \%} & \textbf{Precision \%} & \textbf{F1-Score \%} \\ 
		\midrule
		\multirow{3}{*}{ISCXVPN2016} &  Encapsulation & 98.09 ($\pm$ 1.11) & 98.11 ($\pm$ 1.09) & 98.09 ($\pm$ 1.11) \\ 
		& Service Types & 80.29 ($\pm$ 0.52) & 80.35 ($\pm$ 1.95) & 98.67 ($\pm$ 0.52) \\ 
		& Application Types & 77.16 ($\pm$ 0.44) & 77.99 ($\pm$ 1.49) & 98.67 ($\pm$ 0.69) \\ 
		\midrule
		ISCXTor2016 & Tor Usage & 99.89 ($\pm$ 0.07) & 99.89 ($\pm$ 0.07) & 99.89 ($\pm$ 0.08) \\  
		
		CIC-DoHBrw-2020 & Malicious DoH & 99.96 ($\pm$ 0.02) & 99.97 ($\pm$ 0.02) & 99.97 ($\pm$ 0.02) \\ 
		
		\midrule
		\multirow{2}{*}{USTC-TFC2016} & Benign/Malicious & 99.98 ($\pm$ 0.03) & 99.98 ($\pm$ 0.03) & 99.98 ($\pm$ 0.03) \\ 
		
		& Specific Tools & 99.91 ($\pm$  0.03) & 99.91 ($\pm$  0.03) & 99.91 ($\pm$  0.03) \\  
		\midrule 
		IPTAS-Tbps & Application Types & 88.52 ($\pm$ 3.99) &  89.02 ($\pm$ 2.89) & 88.42 ($\pm$ 3.42) \\ 
		\midrule
	\end{tabular}}
	\label{tab:performance_metrics}
	\caption*{\scriptsize \hfill * All results are presented as mean $\pm 1/2 $ range}
	\end{table}

	For each independent expert model shown in the figure, we conducted ten repeated experiments from the dataset partitioning stage to the end of model training, validating the entire process. The figure displays the mean and range of the model's loss function values and accuracy over the ten training runs. However, due to our oversight, we did not record the validation set results for each epoch when training the Encapsulation and Service type classification using the ISCX-VPN-2016 dataset. Nonetheless, we have recorded and presented the test results for all tasks across all datasets on the test set in Table 4.

\end{document}